\documentclass[12pt,preprint]{aastex}
\usepackage[section] {placeins}
\usepackage{url}

\newcommand {\apgt} {\ {\raise-.5ex\hbox{$\buildrel>\over\sim$}}\ }
\newcommand {\aplt} {\ {\raise-.5ex\hbox{$\buildrel<\over\sim$}}\ } 

\newcommand{\Si}{SMM\,J02399$-$0136}
\newcommand{\Sii}{SMM\,J04431+0210}
\newcommand{\Siii}{SMM\,J123549.44+621536.8}
\newcommand{\Siv}{SMM\,J123707+621410}
\newcommand{\Sv}{SMM\,J14011+0252}
\newcommand{\Svi}{SMM\,J16359+6612}
\newcommand{\Svii}{SMM\,J16366+4105}
\newcommand{\Sviii}{SMM\,J16368+4057}
\newcommand{\Six}{SMM\,J16371+4053}
\newcommand{\Sx}{MM\,J154127+6616}
\newcommand{\asec}{$^{\prime\prime}$}
\newcommand{\OII}{$\left[ \rm{O \ II} \right]$\,}

\newcommand{\nic}{NIC2}
\newcommand{\wfc}{WFC3}

\shorttitle{}

\shortauthors{Aguirre et al.}

\begin{document}

\title{HIGH-RESOLUTION NEAR-INFRARED IMAGING OF SUBMILLIMETER GALAXIES}

\author{Paula Aguirre\altaffilmark{1,2}, Andrew J. Baker\altaffilmark{3}, 
Felipe Menanteau\altaffilmark{3}, \\ 
Dieter Lutz\altaffilmark{4}
and Linda J. Tacconi\altaffilmark{4}
}

\altaffiltext{1}{Departamento de 
Astronom\'ia, Pontificia Universidad Cat\'olica de Chile, Santiago, Chile; paguirre@astro.puc.cl}

\altaffiltext{2}{Departamento de Ciencias F\'isicas, Facultad de Ciencias 
Exactas, Universidad Andr\'es Bello, Santiago, Chile}

\altaffiltext{3}{Department of Physics and Astronomy, Rutgers, The State 
University of New Jersey, 136 Frelinghuysen Road, Piscataway, NJ 08854-8019, USA; 
ajbaker@physics.rutgers.edu, felipe@physics.rutgers.edu}

\altaffiltext{4}{Max-Planck-Institut f\"ur Extraterrestrische Physik, Postfach
1312, D-85741 Garching, Germany; lutz@mpe.mpg.de, linda@mpe.mpg.de}

\begin{abstract}
We present F110W ($\sim J$) and F160W ($\sim H$) observations of ten 
submillimeter galaxies (SMGs) obtained with the {\it Hubble Space Telescope}'s ($HST$'s) 
NICMOS camera. Our targets have optical redshifts in the range $2.20\leq z\leq2.81$ confirmed by millimeter CO or mid-IR spectroscopy, guaranteeing that the two bands sample the rest-frame 
optical with the Balmer break falling between them. Eight of ten are 
detected in both bands, while two are detected in F160W only.  We study their F160W morphologies, applying a maximum-deblending detection algorithm to 
distinguish multiple- from single-component configurations, leading to 
reassessments for several objects. Based on our NICMOS imaging and/or previous dynamical evidence we identify five SMGs as multiple sources, which we interpret as merging systems. Additionally, we calculate morphological parameters asymmetry ($A$) and Gini coefficient ($G$); thanks to our sample's limited redshift range 
we recover the trend that multiple-component, merger-like 
morphologies are reflected in  higher asymmetries.  We analyze the stellar populations of nine objects with F110W/F160W photometry, using archival $HST$ optical data when available. For multiple systems, we are able to model the individual \emph{components} that build up an SMG.
With the available data we cannot discriminate among star formation histories, but we constrain 
stellar masses and mass ratios for merger-like SMG systems, obtaining a mean $\log(M_*/M_{\odot})=10.9\pm0.2$ for our full sample, with 
individual values $\log(M_*/M_{\odot})\sim9.6 - 11.8$.  The 
morphologies and mass ratios of the least and most massive systems 
match the predictions of the major-merger and cold accretion SMG 
formation scenarios, respectively, suggesting that both channels 
may have a role in the population's origin.
\end{abstract}


\section{Introduction} \label{s-intro}

Hierarchical scenarios for structure formation posit that the evolving
mass function of galaxies is driven by the evolving mass function of dark
matter halos \citep[e.g.,][]{mo02} at all redshifts.  In this paradigm,
increasingly large numbers of increasingly massive galaxies will form
through a mixture of ``wet'' (gas-rich) and ``dry'' (gas-poor) mergers,
along with the accretion of intergalactic gas. The simplest version
of this picture would invoke a direct proportionality between dark
and luminous matter, as might be suggested by the observed correlation
between clustering of high-redshift UV-selected galaxies and rest-frame
UV luminosity, which is roughly proportional to galaxy mass (\citealt{ouch04,adel05,hilde09}). However, results from halo occupation
models now indicate that the stellar-to-halo mass (SHM) relationship is
more complex; it has a characteristic peak at a halo mass
$\sim10^{12}M_{\odot}$, and declines toward both smaller and larger
mass, more steeply in the former case (\citealt{somer09,most10,behr10,yang12}). These variations in the SHM
relationship are explained by the interplay of feedback from
supernovae, which is most effective in low-mass halos, and from active
galactic nuclei (AGNs), which becomes more efficient in high-mass halos
(e.g., \citealt{shank06, somer08}). Beyond the form of the SHM relationship, it has also been shown that the peak in star formation
efficiency shifts to higher halo masses at higher redshift (e.g.,
\citealt{wang12, wake11,most13}), a result that adds to several lines
of mounting evidence suggesting that stellar mass build-up happened
much earlier in the \emph{most} massive rather than in the least massive galaxies (e.g., \citealt{cowi96,marc04,thom05,cima06,bund06,kann09}), a
trend known as ``cosmic downsizing.''

Bright submillimeter galaxies \citep[SMGs;][and references therein]{blai02}
represent a key population in tracing the cosmic history of mass 
assembly.  Their high ($\sim 10^{13}\,L_\odot$) luminosities appear to be 
powered mainly by star formation, based on the relative faintness of their 
X-ray counterparts \citep{alex05} as well as their mid-IR spectra \citep{lutz05,vali07,mene07,pope08}, while detections of large molecular gas 
reservoirs demonstrate that SMGs are {\it massive} ($\geq 5 -10 \times 
10^{10}\,M_\odot$) as well as luminous \citep{fray98,fray99,genz03,neri03,grev05,tacc06}.  Bright ($S_{850\,{\rm \mu 
m}} > 5\,{\rm mJy}$) SMGs  at $z\sim2$ have all the properties expected for the 
progenitors of the most massive local spheroids, caught at a crucial juncture 
in their assembly (e.g., \citealt{smai02,swin04,swin06,targ11}).

Explaining the origin(s) of these high star formation 
rates and masses has proved to be more difficult.  SMGs' large, complex 
rest-UV and radio morphologies \citep{chap03,chap04,enge10} provide some
evidence
that their intense starbursts have been triggered by mergers.  Semi-analytic models explaining SMGs' observed number counts suggest that short 
of a top-heavy initial mass function \citep[IMF;][]{baug05}, we {\it need}
multiple early-stage mergers to account for some fraction of the population
\citep{hayw11}.  In contrast, given recent theoretical predictions
that accretion of cold gas along filaments can hasten the coalescence of 
baryons within massive haloes at high redshift \citep{deke06,kere05,kere09}, 
\citet{dave10} have proposed that cold flows rather than mergers can account 
for at least some fraction of the SMG population. In this scenario, SMGs would correspond to super-sized, very massive ($\sim10^{11} -10^{12}\,M_{\odot}$) versions of ordinary star-forming galaxies, fed with gas from minor mergers or smooth cold accretion at rates comparable to their  star formation rates, which after quenching its star formation by some mechanism(s) evolve into systems like the brightest group ellipticals seen today.

Distinguishing between the merger and cold flow scenarios for SMGs' 
striking properties is complicated in two respects by their extreme 
obscuration at short (i.e., rest-UV) wavelengths.  First, high extinction
makes SMGs rather difficult targets for redshift determination; successful
efforts have required the use of radio maps and blind optical spectroscopy
that can be confirmed with CO detections \citep{chap03,chap05,neri03,grev05},
of mid-IR spectroscopy of polycyclic aromatic hydrocarbon (PAH) features \citep{lutz05,vali07,mene07,pope08},
or of CO spectroscopy with dedicated, ultrawide-bandwidth instruments 
\citep[e.g.,][]{swin10a,fray11,lupu11,harris12}.  Such challenges have made it 
difficult to assemble large and homogeneous SMG samples for statistical
analyses.  Second, patchy dust screens have made it difficult to interpret
the morphologies and masses of SMGs' stars in the rest-frame UV.  Initial efforts have 
concluded that SMGs have large stellar masses to match their gas 
masses \citep{smai04,hain11}, but that they are not more likely to appear as major mergers in the rest-frame UV/optical than more typical, lower-luminosity galaxies at the same epoch \citep{swin10b}.

The physical properties of SMGs' stellar populations can be better examined in the rest-frame optical, where dust obscuration is less dramatic than in the rest-frame UV. Hence, we have undertaken a 
program of high-resolution, rest-frame optical imaging of an SMG sample 
in a well-defined redshift range using the {\it 
Hubble Space Telescope} ($HST$).  In this paper, we exploit $HST$'s exceptional resolution to spatially resolve SMGs, characterize their morphologies, identify multiple-component systems that may be interpreted as galaxy mergers, and study the stellar populations in individual merging \emph{components} as well as in single SMGs. We combine our new near-infrared photometry with archival $HST$ optical data to derive stellar population synthesis (SPS) models and estimate stellar masses for each component in single and multiple SMG systems, then computing the mass ratios for merger-like systems. In this way, and to the extent that the merger scenario is correct, our goal is to advance our understanding of the progenitors that coalesce to form a massive SMG. 

We explain the selection of our sample in Section 2. Section 3 describes the acquisition and reduction of our near-infrared observations, and of complementary optical $HST$ data that we have retrieved from the archive in order to do a more complete modeling of our SMGs' stellar populations. In Section 4 we describe the method we use to distinguish multiple-component SMG systems and measure individual magnitudes for each component, and we report the results of this analysis for each target. We also measure several structural parameters and investigate their correlation with the existence of multiple or single components in SMGs; this analysis is presented in Section 5.  In Section 6, we describe our SPS modeling of SMGs and SMG components,  and in Section 7  we discuss our measured stellar masses. Our conclusions are summarized in Section 8. Throughout this paper we use a flat $\Omega_{\Lambda} = 0.7$ cosmology with $H_0 = 70\,{\rm km\,s^{-1}\,Mpc^{-1}}$, and a Chabrier (2003) IMF.

\section{Sample}

We selected  a sample of 10 SMGs for high-resolution near-infrared imaging with $HST$ Near Infrared Camera and Multi-Object Spectrometer (NICMOS) in both F110W ($\sim J$) and F160W ($\sim H$) with the aim of studying their rest-frame optical morphologies, modeling their stellar populations, and estimating their stellar masses. The targets were chosen in 2007 January  and already at that time had optical redshifts confirmed by the published 
detection of a millimeter CO line (\citealt{fray98,fray99,neri03,shet04,grev05,tacc06}) and/or a 
mid-infrared PAH feature (\citealt{mene07,vali07}), both tracers that are closely linked to the far-IR peak of a galaxy's 
bolometric emission. 
We also required that each target's (robust) redshift lay in the range $1.75 \leq z \leq 
3$, placing its $4000\,{\rm \AA}$ break between the centers of the F110W and 
F160W filters, so that we could exploit our two-band 
imaging to constrain each target's stellar population and extinction  
in a spatially resolved sense. Finally, since we desired reliable measurements of F110W$-$F160W ($\sim J-H$) colors for constraining spatially resolved stellar populations, each target had to be observed and detected in both $J$ and $K$ when imaged from the ground \citep{ivis98,bert00,smai02,smai04,fray03,bory05,taka06}, so as to ensure high signal-to-noise ratio (S/N) detections in both $J$ and $H$ bands. The final sample of SMGs and their redshifts are listed in Table \ref{tab:obs}. Two of our targets (\Sviii\ and \Six) already had F160W data obtained by a previous $HST$ program (PID: 9856, PI : Chapman),  so we only observed them in band F110W. 

\section{Observations and data reduction} \label{s-obs}

\subsection{Near-infrared Data}

Our selected SMGs were observed  in the near-infrared F110W $(\sim J)$ and 
F160W $(\sim H)$ bands using the NICMOS  \citep[NICMOS;][]{thom98} and Wide Field Camera 3 
\citep[WFC3;][]{kimb08} on the \emph{Hubble Space Telescope} (PID = 11143, 
PI = Baker). As detailed in Table \ref{tab:obs}, F110W images were obtained 
using the NIC2 camera, which provides a $19\farcs2\times19\farcs2$ field of 
view and a plate scale of $0\farcs075$ pixel $^{-1}$.  In the case of the F160W data, 
six targets were observed with NIC2 and two with WFC3/IR, which has a 136\asec 
$\times$123\asec\,  field of view and $0\farcs13$, pixels. For the remaining pair, 
we retrieved existing NIC2 archival data. All new observations took place 
between 2007 October  and 2009 December and used the \texttt{MULTIACCUM} 
readout mode and a four-point dither pattern to optimize cosmic ray removal.

Our NICMOS F110W and F160W observations were processed using the \texttt{STSDAS} IRAF package \citep{bush94}. Each exposure was calibrated using the routine \texttt{calnica} and corrected for the ``pedestal'' and ``erratic middle column'' effects \citep{bush97, that09}; mosaicking of the dithered datasets was performed using  \texttt{calnicb} \citep{bush97}.

In the case of WFC3 data, calibration of raw files was done with the IRAF \texttt{calwf3} pipeline \citep{quij09}. The 
resulting calibrated images showed a multiplicative offset in the signal level of each detector quadrant, which we rectified by multiplying the lower left, upper left, upper right, and lower right portions of each image by factors of 0.992, 1.004, 0.987, and 1.017 respectively \citep{petro09}. The dithered, calibrated exposures were finally  mosaicked using \texttt{MultiDrizzle} \citep{koek02}, with parameter \texttt{pix$_-$frac}=1 and an output pixel size of $0\farcs075$ pixel$^{-1}$ in order to match the NIC2 resolution.

\subsection{Optical Data}

For better modeling of the stellar populations in our SMG sample, we supplemented our new near-infrared observations with archival $HST$  optical imaging of our targets. Some of our targets have also been observed with ground-based telescopes, but to study the stellar populations in components of SMGs we require the high spatial resolution provided by $HST$.  We compiled optical observations 
previously obtained with the Advanced Camera for Surveys \citep[ACS;][]{ford98}
for the targets listed in Table \ref{tab:optical}, which presents the exposure times and details of each retrieved dataset. All ACS imaging used the Wide Field Channel (WFC), with a 
field of view of $202\farcs\times202\farcs$, and plate scale $0\farcs05$
pixel$^{-1}$.  The raw images were calibrated  with the STSDAS \texttt{calacs} routine and dithered frames were combined using \texttt{MultiDrizzle} with the same parameters as for the WFC3 data. For 
\Siv, however, we downloaded the reduced images and weight maps from the 
Great Observatories Origins Deep Survey (GOODS) $HST$/ACS Treasury Program 
v2.0 release \citep{giav04}; these had a pixel scale of $0\farcs30$ pixel$^{-1}$, 
which we drizzled to match the $0\farcs075$ pixel$^{-1}$ resolution of our NICMOS 
images.

We aligned all available infrared and optical images for each target to the F160W frame, where the S/N is highest. We measured centroids for the few objects we could identify in each target's image set, calculated
the average \emph{x} and \emph{y} displacements relative to the F160W image, 
and shifted each image accordingly. No image rotations were required. In general, all ACS images were aligned 
to within $\sim$ 1 pixel ($0\farcs075$) of each other, but there were shifts 
relative to the F160W band ranging from $\sim1\farcs1$ to $\sim0\farcs3$ in 
R.A. and decl. The F110W images have average R.A.  and decl. 
shifts of $0\farcs3$ and $0\farcs15$, respectively.

The described data reduction process produced infrared and optical images 
that appear flat and free of cosmetic defects and have low background noise. 
All the final  near-infrared and optical images are presented in Figure \ref{fig:stamps}, and discussed in detail in Section \ref{sec:sources}.

\subsection{Astrometric Accuracy}

To estimate the astrometric accuracy of our infrared images, we checked the 
positions of all known objects in our targeted fields. For the NICMOS data, we identified five and six sources from the Hubble Deep Field North near-infrared ($H,K^{\prime}$) catalog of \citet{capa04} in the fields  of \Siii\ and \Siv, respectively. This catalog has an astrometric accuracy of 
$0\farcs03$, and our mean position error relative to its coordinates is 
$0\farcs37$, which gives a total estimated positional error of $0\farcs40$. For the WFC3 images, we matched two sources from the Sloan Digital 
Sky Survey  (SDSS) Data Release 6 in the field of \Six\ and obtained a mean 
astrometric error of $0\farcs30$, which combined with a $0\farcs1$ uncertainty in the SDSS coordinates \citep{pier03} gives  a total estimated uncertainty of $0\farcs40$ for the WFC3 astrometry as well.

\subsection{PSF Matching}

For accurate color calculations, all images were convolved to match the 
widest point-spread function (PSF), which defined our lowest resolution and 
corresponded to the F160W image for those images observed with NICMOS in both 
infrared bands, and to the F850LP image for targets observed with WFC3. We 
used the Tiny Tim web-based application to produce model PSFs for each band 
and for each target's specific position on the detector, and calculated the 
transformation kernels between them and the reference PSF with IRAF's 
\texttt{psfmatch}. We then  applied the transformation kernels to the aligned 
science images and performed a Fourier convolution using \texttt{fconvolve}, which multiplies the Fourier transforms of the input array and kernel and then takes the inverse transform to return a real-space image.\
This method worked successfully for all targets except for the very bright point source in \Si, for which we instead approximated the PSF as a simple two-dimensional Gaussian model with the same full-width at half-maximum (FWHM) as the Tiny Tim model, and convolved to the F160W Gaussian model 
following the procedure described above.

\section{Image Analysis}

\subsection{Detection and Photometry}\label{sec:detection}

Using the known coordinates of our targets, we ran SExtractor \citep{bert96}
in \texttt{ASSOC} mode to select the pixels belonging to each source and measure magnitudes and S/Ns. Since the 4000\,\AA\ break falls between our infrared filters, all targets are brighter in F160W than in F110W, so we chose the former as the  reference for detection and used SExtractor's output segmentation images to measure magnitudes in all remaining infrared and optical bands.

To identify substructure or multiple components in our SMGs and study the stellar properties of each component, we need to disentangle their light profiles and define separate photometry apertures for each individual element. To this end, we ran SExtractor with a $\sigma=2.5$ threshold  and \emph{maximum} deblending  (\texttt{deblend$_-$mincont=0}), so that local peaks in the light 
profile were distinguished as separate objects and we could pinpoint the 
position and extension of each component. The final photometry apertures 
used for each object are indicated in the stamps presented in Figure \ref{fig:stamps}.

In addition to visual and photometric identification of SMG components, we were also interested in measuring morphological parameters and assessing possible correlations with our single versus multiple-system classifications. In this context, we aim to extract the whole galaxy as a single 
object, so that possible substructures or clumps are included in the 
calculated morphological parameters. In addition, for coherent comparison 
between bands we must select the same area in each frame, so that any 
variations with wavelength reflect true morphological differences and not 
selection biases. For these reasons, we ran SExtractor with a lower detection threshold ($\sigma=2.0$) and minimum deblending (\texttt{deblend$_-$mincont=1.0}) on all F160W images, and used the resulting segmentation in the calculations described in Section \ref{sec:morph}.

For each individual object or component, the integrated flux in each filter was calculated as the sum of all pixel values inside the SExtractor segmentation region, and the local sky mean value ($\mu_{\rm{sky}}$) and dispersion ($\sigma_{\rm{sky}}$) were estimated over a $0\farcs5$ wide annulus centered on the target position and located at a radius between $1\farcs5$ and $2\farcs5$ so as to avoid nearby sources. The sky-corrected flux was then calculated as $F_{\rm{corrected}}=F_{\rm{total}}-\mu_{\rm{sky}} N_{\rm{pix}}$, where $N_{\rm{pix}}$ is the number of selected pixels. Uncertainties in the measured flux correspond to the sky noise, corrected so as to take into account the difference in areas over which the sky noise and the total flux were measured. The final formula used was $\sigma_{\rm{flux}}=\sigma_{\rm sky}\sqrt{N_{\rm pix}(1+N_{\rm pix} /N_{\rm sky})}$ where $\sigma_{\rm{sky}}$ is the standard deviation among pixels inside the blank annulus used to estimate the local sky value, $N_{\rm{sky}}$ is the number of pixels in the sky annulus, and $N_{\rm{pix}}$ is the number of pixels  over which the total flux was summed. For WFC3 and ACS images, we also considered the correlated noise introduced by the drizzling process due to the combination of images with partial pixel overlap, and included it in our error estimates following the prescription in Appendix 6 of \citet{case00}.

\subsection{Results for Individual Sources} \label{sec:sources}

In the following paragraphs we describe our near-infrared observations for each target, and discuss our results in the context of previous work in the literature, taking into account as well the optical data we have retrieved from the $HST$ archive. We give special attention to those systems where we have extracted two or more components, which we generally interpret as the building blocks that will merge to form an SMG. Near-infrared and optical magnitudes measured for each object are reported in Table \ref{tab:colors}, including corrections for lensing magnification when appropriate, and in Figure \ref{fig:stamps} we present multi-band $HST$ imaging for each target.

\subsubsection{Multi-component SMGs}

\paragraph{\Si} is the brightest source ($S(850\mu\rm{m})=23.0$\ mJy; \citealt{smai02})
detected in the SCUBA Lens Survey \citep{smai97}, and was 
identified by \citet{ivis98} as a hyperluminous galaxy at $z=2.803\pm0.003$ 
lensed by the foreground cluster A370.  Detection of CO(3--2) emission by \citet{fray98} revealed a large reservoir of molecular gas, indicating a star-formation origin for the high rest-frame infrared luminosity. Optical imaging and spectroscopy by \citet{ivis98} revealed a double counterpart formed by a bright, compact component (L1) hosting a narrow-line AGN (\citealt{vill99,vern01}), and a and a fainter, diffuse companion (L2) associated with Ly\,$\alpha$ and H$\alpha$ emission indicative of a strong starburst.

\citet{ivis10} combined various of the above datasets with new CO(1--0) 
mapping with the Jansky Very Large Array and high-resolution optical 
and infrared imaging to  study the mass and distribution of the 
stars, gas, and dust in \Si. Their  CO(1--0) map revealed a $\sim10^{11}\, M_{\odot}$ 
reservoir of cold molecular gas extended over 25\,kpc in the 
source plane, encompassing sources L1 and L2, and deep 
multiwavelength imaging led to the identification of two additional components 
located to the north of L1 and to the southwest of L2, respectively named L1N 
and L2SW \citep[see Figure 1 in][]{ivis10}. The former was identified as a bright, compact 
source located $\sim 8$\,kpc north of L1, while the latter corresponded to a red feature 
that lies southeast of L1 and southwest of L2, emitting strongly in the IRAC 
bands. The 
total gravitational magnification derived by \citet{ivis10} at the position 
of \Si \ is $2.38 \pm 0.08$, which we have used to correct all source-plane
quantities that are reported here.

The identification of L1N and especially L2SW by \citet{ivis10} relied 
strongly on the NICMOS imaging obtained through our observing program, which 
we reuse here in combination with ACS optical imaging to study in greater 
detail the morphology and stellar population of each component. We have 
reexamined these data and find that the emission associated with L2SW is 
visible not only to the southeast of L1, but also to its northwest, reaching 
out approximately $1\farcs3$ in this direction. We therefore conclude that L1 
is not separate from but actually overlaps with a starburst component that 
extends along $\sim3\farcs8$in the SE-NW direction, and that appears bright 
and continuous in the F160W image but is more fragmented and clumpy in F110W 
and in the optical bands, indicating structured obscuration. As indicated in 
Figure \ref{fig:stamps}, we refer to this component as L1sb through the 
rest of this work. We measured its magnitude in the different bands excluding 
the contribution of the AGN component located in L1, which was modeled as a 
point source and removed from all of the images. The point source model was 
generated using GALFIT \citep{peng02} on the F475W image, where the emission 
is dominated by the AGN; for each of the (PSF-matched) remaining bands we scaled the 
model according to the flux measured inside a circular aperture of the same 
size as the image's PSF FWHM, and subtracted the result from 
the original image. This procedure removed most of the compact emission at 
the position of L1 from the optical F475W and F675W images, leaving only $\sim4\%$ residuals, which could be due 
to small deviations from the assumed roundness of the source. However, for the 
infrared F110W and F160W data, and to a lesser extent for F814W,  there is 
significant residual emission at the position of L1, even after point source 
subtraction. This emission extends over a radius of $\sim0\farcs3$ in all 
three bands and has a slightly irregular structure that is most noticeable in F814W; we attribute it to a strong starburst immediately surrounding the AGN.

 The final apertures used for each component are plotted in Figure \ref{fig:stamps}, which presents the 
point-source-subtracted images. The apparent magnitudes are presented in Table 
\ref{tab:colors}, including the correction for gravitational magnification.  

\paragraph{\Sii}  was first detected by the SCUBA Lens Survey \citep{smai97} in the field 
of the cluster MS\,0440+0204, close to a bright spiral galaxy at $z=0.18$, and it  associated with the extremely red object N4 \citep{smai99}. \citet{smai02} report a  flux density $S(850 \mu 
\rm{m})=7.2\pm1.5$ \,mJy and a  lens magnification of 4.4, which
we also adopt.  It lies at $z=2.51$ as measured from optical spectroscopy \citep{fray03} and detection of the CO(3--2) line \citep{neri03}, which shows a double-peaked profile suggestive of orbital motion \citep{tacc06}.

We detect \Sii\ in both F110W and F160W images, with a total 
extension of $\sim1\farcs4$  in F160W where the S/N ratio  is highest.  
The exceptional spatial resolution of our data allows us to distinguish for 
the first time two separate components in \Sii, which are aligned in a near 
north-south direction and are labeled as sources A and B in the stamps shown in the third row of Figure 
\ref{fig:stamps}. Component A consists of a bright compact nucleus 
surrounded by an extended envelope of diffuse light that is more extended 
towards the east, while component B appears to be an entirely diffuse, 
irregular structure, and is only marginally detected in the F110W image.

 The only previous indication of the two-component nature of \Sii\ comes from 
the detection of a double-peaked  CO(3--2) profile, which is centered at J2000 
coordinates RA=04:43:7.24 and decl.=+02:10:23.8 and has blue- and red-shifted 
peaks separated by  $0\farcs7\pm0\farcs3$ in a north-south direction  
\citep{tacc06}. The CO(3--2) centroid position matches the location of the 
near-infrared emission within the combined  uncertainties of our registration 
($\pm0\farcs4$) and the coordinates reported by 
\citet[][$\pm0\farcs4$]{tacc06}, and falls roughly between components A and 
B, as indicated by the cross in Figure \ref{fig:stamps}. Furthermore, the separation between the peaks is consistent  with the 
$0\farcs53$ distance that we measure between the centroids of components A and B 
along the declination axis. The detection of a double-peaked profile and the 
decomposition into two spatial components suggest that \Sii\  is an ongoing merger 
system.

\paragraph{\Sv} \label{sec:smmj14011_res}

is an $S(850\mu \rm{m})=14.6\pm1.8$\,mJy source detected by the SCUBA  Lens Survey toward the 
cluster A1835, which causes a moderate amplification factor of 2.75 \citep{smai97}. This SMG is associated with a 1.4\,GHz radio 
counterpart and a pair of optical/near-infrared sources at redshift $z=2.56$ denoted J1 and J2, which show no evidence of AGN activity \citep{ivis00,fabi00}, thus suggesting that the enormous energy output of \Sv\ is caused by an ultraluminous 
starburst \citep{fray99}. High-resolution optical and millimeter imaging revealed a complex morphology for J1 \citep{ivis01} with a series of bright knots and a central concentration (J1c), identified as a foreground 
cluster member that is responsible for a modest additional magnification (factor $\sim3 - 5$; \citealt{smai05}). The likely scenario here is that \Sv\ is a two-component system, in which J1 is a massive starburst with some bright clumps of relatively unobscured star formation induced by a dynamical interaction with companion J2. We adopt a magnification factor of 4.0 in calculating rest-frame quantities for J1, and a magnification of 3.5 for J2. 

Our data for \Sv\ include the highest resolution infrared imaging of the 
galaxy obtained so far, and optical (F850LP) ACS imaging that surpasses in 
resolution the previously analyzed F702W/WFPC2 data. We are now able to study 
the system's near-infrared morphology with the same detail that is possible 
in the optical. As shown in the fourth row of Figure \ref{fig:stamps}, components J1 and J2 are clearly detected in bands F110W, 
F160W, and F850LP, but the substructures in J1 are more notable in 
F110W and F850LP, where the knots originally detected by \citet{ivis01} are more clearly defined. Before measuring magnitudes for J1 we remove J1c, which we model in the F850LP image with GALFIT assuming a S\'ersic profile. The best 
fit is a profile with index $n=1.12\pm0.05$ and effective radius 
$R_e=0\farcs36\pm0\farcs02$, which leaves negative residuals of order 10\% of the 
source's flux at the center. These parameters are consistent with the model 
of \citet{nesv07}. The oversubtraction becomes less important as the 
radius increases: it drops to $\sim 5\%$ at a radius of  $0\farcs2$  and 
approaches zero at $\sim0\farcs3$, so the induced error in the photometry of 
J1 is not significant. We then scale this model to the 
remaining bands  by measuring the relative fluxes inside the effective radius 
and subtract the scaled versions. To illustrate our modeling of J1c, in the rightmost panel of Figure \ref{fig:stamps} we show the original image in band F850LP, and in the remaining panels we show the residual images in all filters.  

\paragraph{\Svi} is an intrinsically very faint SMG at $z=2.516$
amplified by the core of the rich cluster A2218, which produces three images denoted A, B, and C \citep{knei04}. The estimated magnification factors for each image are $14\pm2$, $22\pm2$, and $9\pm2$ respectively \citep{knei05}. Optical and near-infrared data reveal a very red counterpart with two blue knots surrounding a red core \citep{knei04}. CO(3--2), CO(4--3), CO(6--5), and CO(7--6) lines show a double-peaked profile, suggesting that the molecular gas traces a rotating disk or torus \citep{shet04,knei05,weis05}.

Our F110W and F160W data for \Svi \ cover the locations of counterparts A and 
B; near-infrared and optical stamps for both images are shown in Figure \ref{fig:stamps}. Counterpart B lies close to the  edge of the NICMOS field in a 
rather noisy section of the image, but near-infrared/optical magnification-corrected magnitudes for sources A and B are consistent within errors (Table \ref{tab:colors}). Hence, for the stellar population analysis performed in Section 5 we use only the results obtained for counterpart A; the agreement in magnitudes measured in all bands gives us 
confidence that the analysis of image A is sufficient to understand the 
stellar population properties of the lensed SMG.

\paragraph{\Sviii}

(N2\,850.4; \citealt{scot02}) is a relatively bright unlensed submillimeter source 
($S(850\mu\rm{m})=8.2\pm1.7$\ mJy) associated with a strong ($S(1.4\,\rm{GHz})=220\,\mu\rm{Jy}$), compact radio source \citep{ivis02}. UV, optical, and millimeter spectroscopy provide evidence that \Sviii\ is a massive, late-stage merger of at least two
components, which has triggered a strong, obscured starburst and an AGN 
\citep{smai03,neri03,swin05,tacc08}. \citet{swin05} identify three dynamically distinct components (denoted A, B, and C) and detect broad-line emission indicative of nuclear emission at the location of B and C,  but do not reach a firm conclusion 
on which of them hosts the AGN. 

We combine new F110W imaging of \Sviii\  with the ACS F814W and NICMOS F160W data published by \citet{swin05}
to study the properties of the starburst and AGN that  likely power this 
source. Component A is not clearly distinguishable from B based on the optical/near-infrared imaging only, so we focus our analysis on B and C, shown in Figure \ref{fig:stamps}, for which we define segmentation photometry apertures that are consistent with 
the  H$\alpha$ velocity field contours in \citet{swin05}. The 
magnitudes measured for sources B and C are given in Table \ref{tab:colors}. 
The F160W emission extends towards the north beyond the boundaries defined 
for components B and C, and traces an irregular tail that may be the result of 
tidal interactions. Component C  becomes fainter as we move from the 
rest-frame optical to UV, but thanks to the addition of the F110W band we 
detect in component B a compact knot that is brighter in F814W than in F110W, 
and therefore emits more strongly in the rest-frame UV. This nucleus suggests 
that the AGN is most likely located in component B, implying that 
the emission in component C is caused by star formation, and that 
the broad optical lines  detected by \citet{swin05} in C are due to  
scattered emission \citep{smai03}.

We have removed component B's point source in order to isolate and study the properties of the underlying starburst. As for \Si, we use  GALFIT to generate the optical model from the F814W image, and then scale according 
to the central flux before subtracting from the infrared images. 

\subsubsection{Single-component SMGs}

\paragraph{\Siii}

 was included in the optical spectroscopic survey of \citet{chap05}, and it is been since found that it hosts a heavily obscured AGN component \citep{alex05,taka06}, with double-peaked rest-frame optical emission lines that likely  indicate a merger geometry or 
rotation along the slit. High-resolution millimeter imaging by \citet{tacc06,tacc08} 
revealed CO(3--2) and CO(5--6) lines with double-peaked profiles and gas kinematics that are interpreted as signs of a compact 
rotating merger remnant.

We detect \Siii\  in both F110W and F160W; it appears as a single central 
nucleus surrounded by a more irregular diffuse component.  There are no 
signs of multiple nuclei, which suggests that the double-peaked profile seen 
in various spectral lines is likely caused by rotation rather than merging 
components.  The F160W emission compact core has a diameter of $\sim 1\farcs$, and the faint envelope has 
an extension of $\sim 1\farcs5$. 

\paragraph{\Siv}

is an unlensed $S(850\,{\rm \mu m}) = 4.7 \pm 1.5$\,mJy source  in the 
Hubble Deep Field North \citep{chap04,chap05}, which has been resolved through high-resolution millimeter continuum imaging 
\citep{tacc06, tacc08} into a double source with a separation of $\sim2\farcs5$. One CO peak is coincident with a radio/optical/near-infrared source (\citealt{swin04,tacc08}), while the second is only detected strongly in the radio continuum, with additional radio blobs seen around the (sub)millimeter source. These observations suggest a complex environment with several potential interacting sources, but we focus our analysis on the only component visible at near-infrared and optical wavelengths, for which we measure magnitudes in F160W and F110W, and in GOODS optical bands  F606W ($V$), 
F775W ($i$), and F850LP ($z$). For band F435W ($B$), we can only estimate a 
magnitude limit. As seen in Figure \ref{fig:stamps}, the source is small 
and compact, with a  half-light radius of  $\sim 3$\,kpc. \citet{swin04} distinguish in the optical ($V,i,z$) bands two nuclei 
separated by $\sim0\farcs2$ that coincide with knots of 
H$\alpha$ emission, and that are therefore interpreted as two separate objects, but our data show that the F160W emission envelops both components, so they are probably just 
emission peaks in a single object with high but irregular absorption.
  
\paragraph{\Sx}

is an unlensed source detected in a deep 1.2\,mm MAMBO survey of A2125, for which \citet{bert00} identified a radio and $K$-band counterpart, plus a faint arc seen in $J$ and $K$ imaging that appears to connect \Sx\  with a brighter source located $\sim 4\farcs$ to the southeast. \Sx \ is not detected in our F110W image, so we can only constrain its 
magnitude. Considering a $3 \sigma$ threshold and a 2\asec\ aperture, as in 
the previous limit reported by \citet{bert00}, we find that F110W$>25.7$.
In the F160W band,  we detect an irregular source with magnitude 
F160W=$23.65\pm0.05$. We also see traces of the arc reported by \citet{bert00} in our  F110W 
data and more clearly in the F160W image $-$ starting at a distance of $\sim2\farcs5$ SE of \Sx, beyond the region shown in Figure \ref{fig:stamps} $-$ so we are presumably 
looking at a less-obscured structure at the same redshift.

\paragraph{\Svii}

(N2\,850.2) is an unlensed source detected in the SCUBA 8\,mJy survey of the ELAIS N2 field \citep{scot02,fox02}, with a strong radio counterpart and complex, multicomponent $K$-band emission \citep{ivis02}, but undetected in optical $VRI$ bands and without identifiable AGN features \citep{chap05,swin04}. Millimeter observations have revealed a double-peaked CO(3--2) line \citep{grev05,tacc06} and a large ($\sim 10^{10}\,M_{\odot}$) dynamical mass enclosed within a small (1.6 kpc) radius, which are interpreted as evidence of a compact merger remnant \citep{tacc08}.

In our NICMOS data, we detect a F160W $\sim$25 mag source that is close 
to the edge of the image and therefore has a slightly larger photometric error 
than do our other targets. As shown in Figure \ref{fig:stamps}, this source is compact, with a Petrosian radius of 
2.86\,kpc in F160W, and it matches one of the $K$-band components and the radio source detected by \citet{ivis02}. However, the source is undetected in the F110W and F814W bands, for which we derive 5$\sigma$ AB magnitude lower limits of 26.68 and 28.27, respectively.

\paragraph{\Six} \label{sec:smmj16371_res}

\Six\ is an unlensed source detected in a 1.2\,mm MAMBO survey of the ELAIS-N2  field \citep{grev04} that lies at $z=2.38$ as measured from optical \citep{chap05} and CO \citep{ grev05} emission lines. Near-infrared spectroscopy has revealed H$\beta$ and  \OII\ emission consistent  with the presence of an AGN \citep{taka06}, but detection of mid-infrared PAH features suggest that its  bolometric luminosity is dominated by starburst activity. In this work, \Six\ is detected and resolved in both near-IR bands and in the F775W optical band;  it has a central bright compact nucleus surrounded by a diffuse, irregular component.

\section{MORPHOLOGICAL ANALYSIS}\label{sec:morph}

To study the infrared morphologies of our galaxies, we calculated a set of 
non-parametric measurements  that quantify their internal structure. The Gini coefficient ($G$) gives a quantitative measure of how unequally a galaxy's light is distributed among its pixels \citep{abra03}; it takes a minimum value of zero if the light is distributed uniformly among all the pixels, and a maximum of 1 if all the light is concentrated in one pixel. The concentration index ($C$; \citealt{abra94})  is the ratio of the flux within an inner aperture to the flux outside it and is a proxy for a galaxy's bulge-to-disk ratio, while the asymmetry parameter ($A$) quantifies the degree to which the light of a galaxy is rotationally symmetric \citep{lotz04}. We also studied the 
second-order moment of the brightest pixels, $M_{20}$ \citep{lotz04}, which depends on the spatial distribution of any bright nuclei, bars, or off-center concentrations. These 
parameters have the advantage of not requiring any assumptions about a 
galaxy's light profile, so they can be applied to irregular or 
disturbed galaxies; here we wish to evaluate if they correlate with the presence or absence of multiple components in merger-like SMGs.

 All morphological parameters were computed with the \texttt{PyCA} software 
\citep{menan06}, which implements the mathematical definitions detailed in the Appendix. To determine the  aperture within which all statistics are calculated, 
\texttt{PyCA} takes as input a SExtractor segmentation image used to define 
the total galaxy region; for this purpose, we used the segmentation images 
generated from F160W images with minimum deblending settings. In this way, we can examine if possible 
components or clumps affect the calculated morphological 
parameters, even if they are initially selected within a unified structure, and we 
can make a coherent comparison between the results obtained for F110W and 
F160W data, since we select the same physical area in both filters. Following the analysis by \citet{lotz04} of noise effects on the calculation of $G$, $M_{20}$, and $A$, we require a minimum S/N of 5 to perform a morphological 
analysis, and we assume a systematic uncertainty of $10\%$ in our measurements.

We measured morphological parameters for all targets in bands F110W and F160W, except for 
\Sx\ and \Svii, which were not analyzed in band F110W due to low S/N. For \Svi, we know that both counterparts (A and B) are highly lensed 
and geometrically distorted by the cluster's gravitational potential, so 
parameters that depend on the source's spatial distribution, like $A$, $C$, 
and $M_{20}$, are not meaningful. The individual measurements are presented in 
Table \ref{tab:morph}, and the full set of results is summarized in Figure 
\ref{fig:morph}. 

The sources are generally compact, with a mean half-light  radius of $\langle r_{\rm{h}} \rangle=3.4\pm0.3$ kpc  in the F110W band and $\langle r_{\rm{h}} \rangle=3.6\pm0.2$ kpc in F160W. If we correct for gravitational magnification for those targets that are lensed by cluster potentials, the F110W and F160W source-plane radii are $\langle r_{\rm{h}}\rangle=2.5$ kpc and $ \langle  r_{\rm{h}}\rangle=2.7$ kpc, respectively. These results agree with recent work by \citet{swin10b}, who measure $r_h$, $G$, 
and $A$ in  F160W and in the optical $I$ band  for a sample of 23 SMGs with 
redshifts in the range $z=0.7-3.4$. Their sample includes one of our objects, 
\Sviii,  for which they measure a F160W half-light radius of $3.4 \pm 0.4$\,kpc, in 
excellent agreement with our result for the object's half-light semi-major axis, $a_{\rm{h}}=3.44$ kpc. For the full sample, they obtain median sizes 
of $2.3\pm0.3$ kpc and $2.8\pm0.4$ kpc in the observed optical and near-infrared, 
respectively. \citet{swin10b} also calculate and analyze the morphological 
parameters of \Si\ (L1/L2) for  the F160W band using source-plane $HST$ imaging derived from a gravitational 
lensing model, and obtain $r_{\rm{h}}=2.8\pm 0.4$\,kpc, $G=0.88$, and $A=0.30$.  We measure a lensing-corrected half-light radius of $r_{\rm{h}}=2.52$\,kpc, consistent within errors with their value. \citet{swin10b}  obtain $G=0.83$ and $A=0.66$ for this system, but a direct comparison with our $G$ and $A$ is not conclusive due to possible differences in the regions used by the two studies. 

The median Gini coefficient and asymmetry for our sample are $(G, A)=(0.79
\pm0.17,0.63\pm0.02)$ in F110W, and $(G,A)=(0.70\pm0.11,0.51\pm0.01)$ in 
F160W. These parameters characterize the morphologies of our targets at 
rest wavelengths of 3211$\pm$215\,\AA\ and 4584$\pm$238\,\AA, respectively. 
\citet{swin10b} differ in the median values for their SMG sample and obtain $G=0.56 \pm 0.02$ and $A=0.27\pm 0.03$ in the observed near-infrared, but since calculation of morphological parameters depends strongly on aperture and signal-to-noise, quantitative comparison between different datasets is not straightforward \citep{lisker08}. We can still contrast different interpretations regarding the utility of  morphological  parameters as indicators of merger-like configurations. For example, \citet{swin10b} compared the morphologies of  SMGs with those of UV/optically selected 
star-forming galaxies, and found that the mean sizes, asymmetries, and Gini 
coefficients agree within uncertainties. Their conclusion is that the 
rest-frame UV/optical morphological parameters of SMGs are not more likely to indicate major mergers than they do for more typical galaxies at the same redshifts, even if spectroscopy at other wavelengths does reveal complex kinematics or merging systems.
In our sample, however, we obtain a larger spread in $A$ than \citet{swin10b} and 
find that the values do  correlate to some extent  with the existence of 
multiple nuclei or complex configurations, especially in F160W. In Section 
\ref{sec:sources} we analyzed each source in detail and classified each as a multiple or single-component system based on the segmentation images produced by SExtractor, which thanks to the high spatial resolution of our images is capable of separating independent components separated by a few kiloparsecs. In all cases, the identification of multiple components is supported by previous interferometric evidence,  which consistently reveals complex, merger-like dynamical structures. Thus,  we  classified five of our targets as multiple-component 
systems and five as single-nuclei, as indicated in the last column of Table 
\ref{tab:morph}. In parallel, we calculated morphological parameters  for 
each target, and as shown in Figure \ref{fig:morph}, the only significant trend we find is that sources for which we identify multiple components (plotted as dots) have systematically larger asymmetries than 
single objects (plotted as triangles) in band F160W. The median values are $A_{\rm M} = 0.65\pm0.01$ and $A_{\rm S}=0.38\pm0.01$ for multiple and single-component systems respectively, and are plotted as open symbols in the middle-right panel of Figure \ref{fig:morph} . The 
only outlier from this trend is \Sx, which has a high asymmetry despite being 
included in the single-component group; this object's high value for $A$ is 
explained by its unusual irregular, diffuse appearance. The trend of increased $A$ for merger-like systems does not hold for band F110W, but at shorter 
wavelengths the asymmetry tends to increase for all objects, and the effect of 
multiple components may be masked by the existence of brighter knots of star 
formation or structured obscuration. For $M_{20}$ and $C$, there are no clear 
differences in the measured values between the two source groups in any filter.

We believe that our ability to detect a correlation between $A$ and probable merger-like configurations owes mainly to the homogeneity of our sample. Our targets span a narrow range of redshifts, with 
$z=2.2-2.8$, so observations in band F160W correspond to similar rest-frame 
wavelengths for all objects. However, in the sample of \citet{swin10b}, a 
significant number of objects lie at $z<2$ and there is a much larger spread 
in redshift; F160W imaging consequently reflects a mixture of rest-frame 
morphologies, with 40\% of the sample being observed at longer rest 
wavelengths than the average for our targets. 
By comparing F110W and F160W 
observed morphologies, we find that $G$ and $A$ tend to decrease as we 
move to redder bands, where dust obscuration is less structured. According to 
our analysis, we would expect an object with complex, merging morphology to 
have at least higher asymmetry than a single-component system, but this effect 
could be partially compensated if the multi-component object is actually being 
observed at a redder rest wavelength. Therefore, the $A$ versus 
multiplicity trend seen in our sample could be blurred if the target selection 
is not homogeneous in terms of redshift, which explains the different 
assessments of morphological parameters as indicators of merger likelihood from this work and \citet{swin10b}. Although the \citet{swin10b} sample includes SMGs at lower redshifts, a lower redshift is no guarantee that an object is more evolved that objects at $z=2.2 - 2.8$, so we believe that our differing morphological measurements are not reflecting an evolutionary trend. As a further indication of the difference between samples, only two objects in the \citet{swin10b} sample would definitely have matched our 
selection criteria. However, our targets' mean 
redshift ($\left<z\right>=2.1$) is closer to the peak of the SMG redshift 
distribution \citep{chap05}, so our conclusions may be at least representative for the population.

\section{Stellar masses}

For SMGs with data and significant detections in three or more bands, we can compare the observed magnitudes with those expected for evolved  stellar populations of different ages and star formation histories (SFHs) at the same redshift, and in this way study their physical properties. We used the GALAXEV library of SPS models and the 
\texttt{csp$_-$galaxev} code \citep[][BC2003]{bruz03} to compute the spectra of evolved stellar populations with solar metallicity and a \citet{chab03} IMF at various ages for an assortment of SFHs, including reddening. For the SFH, we adopted an exponentially declining function with an $e$-folding 
timescale $\tau$ in the range 
$\tau=0.1-5$ Gyr to cover a  representative set of models. For each model, we computed the evolved spectra at ages between 0 and 5 Gyr with uniform spacing of 0.1 Gyr, and included the effect of dust attenuation by applying a \citet{calz00} extinction ($A_V$) curve, with $A_{V}=0.001$ ($\sim$ no extinction) to $A_{V}=5$ in steps of $\Delta A_V=0.25$.

With the exceptional spatial resolution provided by $HST$, we can characterize the components in multiple systems \emph{individually}, and thus study the stellar properties of the galaxies that build up SMGs. From our analysis of $HST$ near-infrared imaging, we have identified multiple components in four SMG systems (\Si, SMM \,J04431+0210, \Sv, and \Sviii). There is evidence from previous millimeter interferometry that \Siv\ is also a complex system with several potential interacting components (\citealt{tacc06,tacc08}), but only one is detected in the near-infrared/optical, so for the sake of our stellar population analysis we treat it as a single SMG. For SPS modeling of multiple-component systems, we do not take into account previous ground-based data, since we are specifically interested in studying them at $HST$ resolution, where components can be resolved. However, for analysis of the six remaining SMGs where we identify single stellar components (\Siii, \Siv, \Sx, \Svi, \Svii, and  \Six), we can improve our stellar population modeling by including results from previous authors. In the analysis of \Siii, SMM J123707+621410, \Svii, and \Six, we include IRAC photometry from \citet{hain09}; in the case of \Siii, and \Siv\, we consider as well the optical magnitudes reported in \citet{bory05}. For \Sx, there are no additional measurements in the rest-frame UV/optical.

For each object with multiwavelength photometry, $\chi^2$ minimization can be used to find the best-fitting  stellar population model.  However, for a limited number of photometric points,  there will be a broad set of models with different star formation timescales, ages, and reddenings that provide similarly good fits, as demonstrated for instance in \citet{hain11}. To illustrate this point in the context of our data, we show in Figure \ref{fig:smmj02399_ssp} synthetic color-color plots for our set of stellar population models at the redshift of \Si\ ($z=2.81$), and we also plot the observed photometry for each component in the \Si\ system, as well as associated error ellipses with  semi-major and semi-minor axes $a=\sigma_{1}$ and $b=\sigma_{2}$, for $\sigma_{1}$ and $\sigma_2$  the photometric errors for the two colors in the plot as derived from Table \ref{tab:colors}. We find that for L2 and L1sb there are numerous SPS models that reproduce the observed colors, while for L1N  there are various evolved stellar populations that  simultaneously fit all but the F475W observed magnitude, which we attribute to our simple treatment of dust attenuation. The \texttt{csp$_-$galavex} code applies the two-component model of \citet{charlot00} that accounts for attenuation due to dust in the outer envelopes of the dense clouds where stars are formed and in diffuse cirrus clouds in the interstellar medium. It is likely, however, that the dust distribution in SMGs is patchy and inhomogeneous (\citealt{chap04}), introducing complex reddening effects. Detailed modeling of a clumpy dust geometry would necessarily introduce a large number of free parameters and degeneracies, and thus a significantly larger number of data points would be required for a meaningful result.

 With the current dust treatment, we can find compatible SPS models for all objects in our sample if we relax our matching criteria and require that the predicted magnitudes are within $3\sigma$ of the observed magnitudes in all available bands. The set of compatible SPS models spans wide ranges in age, timescale $\tau$, and extinction $A_V$, so we are therefore unable to reliably estimate these parameters, but we can constraint stellar masses, which are more narrowly distributed, with a typical dispersion $\Delta \log M_*=\pm0.1$. To properly account for the uncertainties due to age, extinction, and SFH degeneracies, we used Hyperzmass \citep{pozz07} to calculate the stellar masses predicted by all selected models, and we adopt the mean and dispersion of the resulting distribution as our final value and uncertainty in $M_*$. With this approach, we estimated stellar masses for all SMGs and SMG components except \Sx, which we detect only in F160W and has no additional optical/near-infrared photometry in the literature. For \Sii, we only have F110W and F160W magnitudes for both components, but since these filters bracket the Balmer/$4000\,{\rm \AA}$ break, they provide an effective constraint on the rest-frame SEDs, so we can still estimate stellar masses for \Sii A and \Sii B, albeit with moderately larger uncertainties. Our results are summarized in Table \ref{tab:mass}, where we report our stellar mass estimates and mass ratios for multi-component systems.

Previous analyses of SMG stellar populations have produced  mean  masses 
that differ by factors up to $\sim10$. \citet{hain11} subtracted the near-IR 
continuum excess contributed by AGNs and obtained a median 
$\log(M_*/M_{\odot})\sim10.8$, while other studies also based on SED fitting 
suggest higher masses in the range $\log(M_*/M_{\odot})=11.4 - 11.8$ 
\citep{bory05,dye08,micha10}. The latter estimates support the scenario 
proposed by \citet{dave10}, who studied the nature of rapidly star-forming 
galaxies at $z=2$ in cosmological hydrodynamic simulations and concluded that 
SMGs may not be scaled up versions of local major mergers seen as ULIRGs, but 
can  be explained instead as super-sized versions of ordinary star-forming 
galaxies fed by infalling gas-rich satellites. This cold accretion model, 
however, implies stellar masses in the range $\log(M_*/M_{\odot}) \sim 11-12$, 
so confirmation of lower masses like those estimated by \citet{hain11} would 
undermine this interpretation.

We have estimated the total stellar masses of nine SMGs, four of which are 
separated into two or more interacting components. If we add up these 
individual concentrations and consider the total mass of each SMG system, we obtain 
values in the range $\log(M_*/M_{\odot})\sim9.6 - 11.8$, with a sample mean of  
$\log(M_*/M_{\odot})=10.9\pm0.2$. We have four objects in common with 
\citet{hain11} and \citet{micha10} (\Siii, \Siv, \Svii, and \Six); for the first 
two of these we estimate masses that are a factor $\sim 4 - 8$ lower than those 
of \citet{micha10} (even after scaling our values to their Salpeter IMF), but within a factor $\sim 2$ of the results of \citet{hain11}. For \Svii\ and \Six,  however, our mass estimates for single-component SMGs are in good agreement with those of \citet{micha10}. In general, we find that our mean stellar mass is consistent with that of \citet{hain11}, but individual masses span the range of values from the more modest 
measurements of \citet{hain11} to the higher predictions of \citet{dave10}, 
consistent with both formation channels $-$ major mergers and cold accretion $-$
playing a role in the birth of SMGs. 

Aside from stellar masses, we can also find indications of these two formation 
mechanisms imprinted in the morphologies of the SMGs in our sample. First, we 
find that half of our sample  presents high asymmetries that correlate with complex, multi-component morphologies that may be attributed to mergers or interacting systems- consistent with previous spectroscopic studies that detect complex dynamical structures. For multiple systems, we are able to 
estimate stellar masses for \emph{each} component and therefore their mass ratios.
We find, for example, that for binary sources like \Sii\ and \Sviii\ with moderate total 
masses ($<10^{11}M_{\odot}$), the proportions are respectively $\sim$3:1 or close to unity, similar to the ratios observed by 
\citet{dasy06} for local ULIRGs.
This result is consistent with that of \citet{enge10}, who measure dynamical mass ratios for four binary SMGs and broadly conclude that most SMGs are major mergers. However, in 
the massive system \Si, the mass ratios that we have now measured between the 
central concentration (L1sb) and its companions (L1N and L2)$-$ not available 
to \citet{enge10}$-$ are significantly larger ($\sim$50:1  and $\sim$126:1, 
respectively), suggesting that the SMG starburst is taking place in a system that may be undergoing a minor merger. We also find that some of the most massive objects like \Sx, \Siv, and \Six\  show no distinct mass concentrations or evidence of merging nuclei, and have optical rest-frame morphologies more consistent with  the simulated rapidly star-forming galaxies at $z=2$ in the work of \citet{dave10}. We conclude that depending on which extreme 
of the SMG mass range we are looking at, there is observational evidence to 
support either the major merger or the cold accretion  scenario, so it is 
entirely possible that we are studying a heterogeneous population in which 
different formation mechanisms are involved.

\section{CONCLUSIONS}

We have presented $HST$ NICMOS and WFC3 near-infrared imaging in 
bands F110W ($\sim J$) and F160W ($\sim H$) for a sample of 10 SMGs with 
redshifts in the range $z=2.2-2.8$, which we have combined with existing optical ACS 
data in order to study their morphologies, stellar populations, and stellar 
masses. Out of our sample, eight targets were detected in both NICMOS bands, and two were detected only in F160W.  The sample was deliberately selected 
so that each object had a CO or PAH-confirmed redshift that put the 
Balmer/4000\,\AA\ break between the F110W and F160W bands; all objects are 
brighter in the latter band, so we used it as the reference detection frame.

To study the possible multi-component, merger-like nature of our targets, we 
took advantage of the maximum deblending option in SExtractor to  assess 
whether each system could be resolved into two or more stellar structures 
coalescing to form a massive SMG. Through this analysis, and with the support of previous dynamical evidence,  we were able to 
classify four of our targets (\Si, \Sii, \Sv, and \Svii) as multi-nuclei 
systems at different merger stages, and measured magnitudes for  
individual components. Previous CO data show that \Siv\ is also 
an early-stage merger system with only one component visible in the rest-frame 
optical \citep{tacc08}.  Thus,  five objects in our sample show 
rest-frame optical and/or molecular gas morphologies that are consistent 
with ongoing mergers.  The high resolution and long wavelength of our {\it 
HST} imaging have allowed us to characterize our targets with greater 
confidence than has been possible in the past: we have newly identified the 
large central mass concentration underlying the bright AGN in \Si, the binary 
merger of \Sii, and the location of the optically bright AGN in \Sviii. With the advent of high-resolution millimeter/submillimeter observations with ALMA, it will be possible to discern merger-like configurations and identify components  in samples selected from the high-redshift end of the SMG population, as shown in the study of a massive starburst at $z=4.7$ by  \citet{wagg12} and \citet{ carilli13}.

In addition to this photometric analysis, we have also measured the mean source 
sizes and studied the morphologies of our SMG sample.  We calculated the asymmetry ($A$), Gini coefficient 
($G$), second-order moment of brightest pixels ($M_{20}$), and concentration 
($C$) in bands F110W and F160W, in order to study possible morphological 
evolution with wavelength, and to determine whether these parameters are 
useful as indicators of mergers. The observed circularized radii are 
$\langle r_h \rangle=3.4\pm0.3$\,kpc for F110W and $\langle r_h\rangle=3.6\pm0.2$\,kpc for F160W; this 
last result and results for individual objects are comparable to  those of earlier studies  
\citep[e.g.,][]{swin10b,targ11}.  We also find that parameters $G$, $C$ and $M_{20}$ 
do not show any clear evolution between bands F110W and F160W, and do not 
correlate with the presence or absence of multiple components. However, $A$ 
is  higher in F110W than in F160W; the mean values are 
$A_{\rm F110W}=0.63\pm0.02$ and $A_{\rm F160W}=0.51\pm0.01$, 
respectively. This trend is consistent with deeper, more structured 
obscuration and more concentrated star activity in bluer bands, and is also in 
agreement with previous measurements by \citet{swin10b}. When we compare the 
mean $A$ for the set of objects for which we identified multiple 
components with the mean values for single-nucleus SMGs, we find that in band 
F160W, $A$ is higher for multiple systems. This trend between 
$A$ and multiplicity indicates that $A$ is a useful 
indicator of merging configurations, with the caveat that the sample under 
analysis must be homogeneously selected in terms of redshift for the trend to avoid being 
blurred out.  Previous work by \citet{swin10b} concluded that the asymmetry of 
SMGs was comparable to that of UV-selected galaxies at the same epoch and 
therefore not a good reflection of their merger-like morphologies, but we 
believe that in this case the expected increase in asymmetry with 
morphological complexity may have been masked by the sample's larger spread in redshift.

We have carried out a SPS analysis for all 
objects detected in at least two $HST$ near-infrared bands, using as well optical $HST$ archive data when available, plus IRAC and/or ground based optical/near-infrared measurements for single-component systems. 
We find that satisfactory SPS models span a wide range of ages and extinctions, so we are not 
able to determine the SFH or a galaxy's dust content 
based on the available data. However, we are able to constrain stellar masses for nine SMGs, and in four multiple systems we obtain stellar masses for each of the components that are likely merging to build up a larger SMG. Determination of SMG stellar masses allows us to assess the predictions of 
the cold accretion scenario \citep{dave10}, which has been proposed as an 
alternative to the typical major-merger formation model. The former predicts 
number densities and clustering that match current observations, but it 
requires stellar masses in the range $\log(M_*/M_{\odot})\sim11 - 12$, and 
would be less preferred if more modest estimates like those of \citet{hain11} are correct. For our sample, we find that if 
we consider  the total combined masses in each SMG system, our estimates range 
from $\log(M_*/M_{\odot})\sim 9.6$ to $\sim11.8$, with a mean of 
$\log(M_*/M_{\odot})=10.9\pm0.2$. These values suggest that both mechanisms 
may be operating, and a more detailed analysis of stellar masses versus
morphologies for individual objects supports this view.  Either the most 
massive systems show no evidence of multiple concentrations or merger-like 
morphologies, resembling the simulated examples of \citet{dave10}, or they 
can be separated into two or more merging components whose high mass ratio 
does not agree with the ``major merger'' definition but more likely suggests 
accretion of small companions. On the other hand, less-massive objects with 
clear merging morphologies do have mass ratios comparable to those observed 
for local ULIRGs, and could therefore be interpreted as scaled-up versions of 
that local population. We speculate then that both the major merger and 
the cold accretion mechanisms have contributed to the formation of the SMG 
population, but on opposite extremes of the observed stellar mass range.

\acknowledgments

The authors thank Galina Soutchkova and Jason Kalirai for their help in 
coordinating the {\it HST} observations, Saquib Ahmed for his 
contributions to the early stages of our analysis of the data, and an anonymous referee for comments that helped improve the paper.  Support for 
this project was provided by NASA grant HST-GO-11143.01-A. P. Aguirre 
acknowledges support  from the FONDAP Center for Astrophysics 15010003, 
BASAL CATA Center for Astrophysics and Associated Technologies, the Chilean 
National Committee for Research in Science and Technology (CONICYT), MECESUP, and Universidad Andres Bello Regular Project DI$-$05$-$11$/$R.

\appendix
\section{CALCULATION OF MORPHOLOGICAL PARAMETERS}

For clarity and ease of comparison with other work, we review here our definitions and practices for measurement of morphological parameters.

\subsection{Gini Coefficient ($G$)}
We calculated $G$ according to the formulas in \citet{glas62}. This definition depends heavily on the 
selection of pixels that belong to a galaxy; we used all pixels inside an elliptical aperture with semi-major axis $R_p=1.5\cdot r_p$, where $r_p$ is the \citet{petr76} radius at which the ratio of the surface brightness measured locally over the surface brightness 
averaged across the enclosed region is equal to a fixed number  $\eta=\frac{\mu(r_p)}{\mu(r<r_p)}$, in this work set to 0.2. The elliptical aperture axial ratio was obtained by dividing the effective major and minor radii measured by SExtractor, and to characterize the source size we measured the \emph{half-light} major ($a_{\rm{h}}$) and minor ($b_{\rm{h}}$) axes, defined as the dimensions of the ellipse that encloses half of the flux inside the Petrosian radius. For a better  representation of the sizes of galaxies that are geometrically distorted by gravitational lensing, we then calculated  the circularized radius $r_{\rm{h}}=\sqrt{a_{\rm{h}}\cdot b_{\rm{h}}}$, which is not dominated by global cluster shear and scales as the square root of the magnification factor $\mu$. The final elliptical apertures were carefully checked to ensure that there was no contamination from other sources in the calculation of morphological parameters.

\subsection{Concentration Parameter ($C$)}
The concentration parameter $C$ is generally defined as  the integrated flux 
inside an inner isophote or radius $\alpha R$ divided by the flux inside $R$, 
for some $\alpha<1$ \citep{abra94}. Following \citet{menan06} we used 
$\alpha=0.3$, so that $C$ is computed as:

\begin{equation}
C=f(0.3R)/f(R) \,\,,
\end{equation}

\noindent with $f(r)=2\pi  \int_0^R I(r)dr$. The distance $R$ was defined as $R=1.5\cdot r_p$.

\subsection{Asymmetry ($A$)}
The asymmetry parameter $A$ measures the fractional difference between the original 
image  and the same image rotated by 180$^{\circ}$ around its centroid, 
corrected for variations in the image background. The pivot point for rotation is 
defined as the position that yields minimum asymmetry \citep{cons00}. To find 
it, we used the galaxy's luminosity centroid as a first guess, calculated $A$, 
and then repeated using as centers the eight surrounding points in a $3\times3$ 
grid, with a grid spacing of 0.1 pixels. If $A$ was minimized at the grid 
center, we chose this point as the rotation center; otherwise the procedure 
was repeated for a new grid centered on the point where the asymmetry was 
lowest, and continued until a minimum was found. 

The effects of correlated noise were removed by performing the same asymmetry 
measurement on a portion of neighboring blank sky. The sky asymmetry was 
minimized and scaled by the size of the object relative to that of the blank region, 
normalized by the object's total flux and subtracted from the value measured 
for the galaxy,  so that the final formula used to compute $A$ resulted in
\begin{equation}
A=\rm {min} \left( \frac{\sum |\emph{I}_0-\emph{I}_{\rm{R}} | }{\sum 
\emph{I}_0} \right ) -\rm min \left( \frac{\sum | \emph{B}_0-\emph{B}_{\rm{R}} 
| }{\sum \emph{I}_0} \right )
\end{equation}
where $I_0$ is the original source image, $B_0$ represents the blank 
sky pixels, and $I_R$ and $B_R$ are the corresponding images rotated by 180$^{\circ}$.

\subsection{Second-order Moment of Light of the Brightest 20\% of the Galaxy's Pixels ($M_{20}$)}

The total second-order moment of light corresponds to the sum of all pixel fluxes 
multiplied by their squared distances to the centroid, such that 
$M_{\rm{tot}}=\sum^n_{i=1} M_i =\sum_{i=1}^n f_i [(x_i-x_c)^2+(y_i-y_c)^2]$;  high values 
reflect the existence of bright nuclei away from the galaxy's centroid or 
extended light distributions like bars or spiral arms (\citealt{lotz04}). Based on this 
definition, $M_{20}$ was calculated as
\begin{equation}
M_{20}=\log_{10} \left( \frac{\sum_i M_i}{M_{\rm{tot}}} \right),  
\qquad {\rm while} \quad \sum_i f_i < 0.2 \cdot f_{\rm tot}
\end{equation}

{}

\clearpage

\begin{figure}
\centering
\includegraphics[scale=0.8]{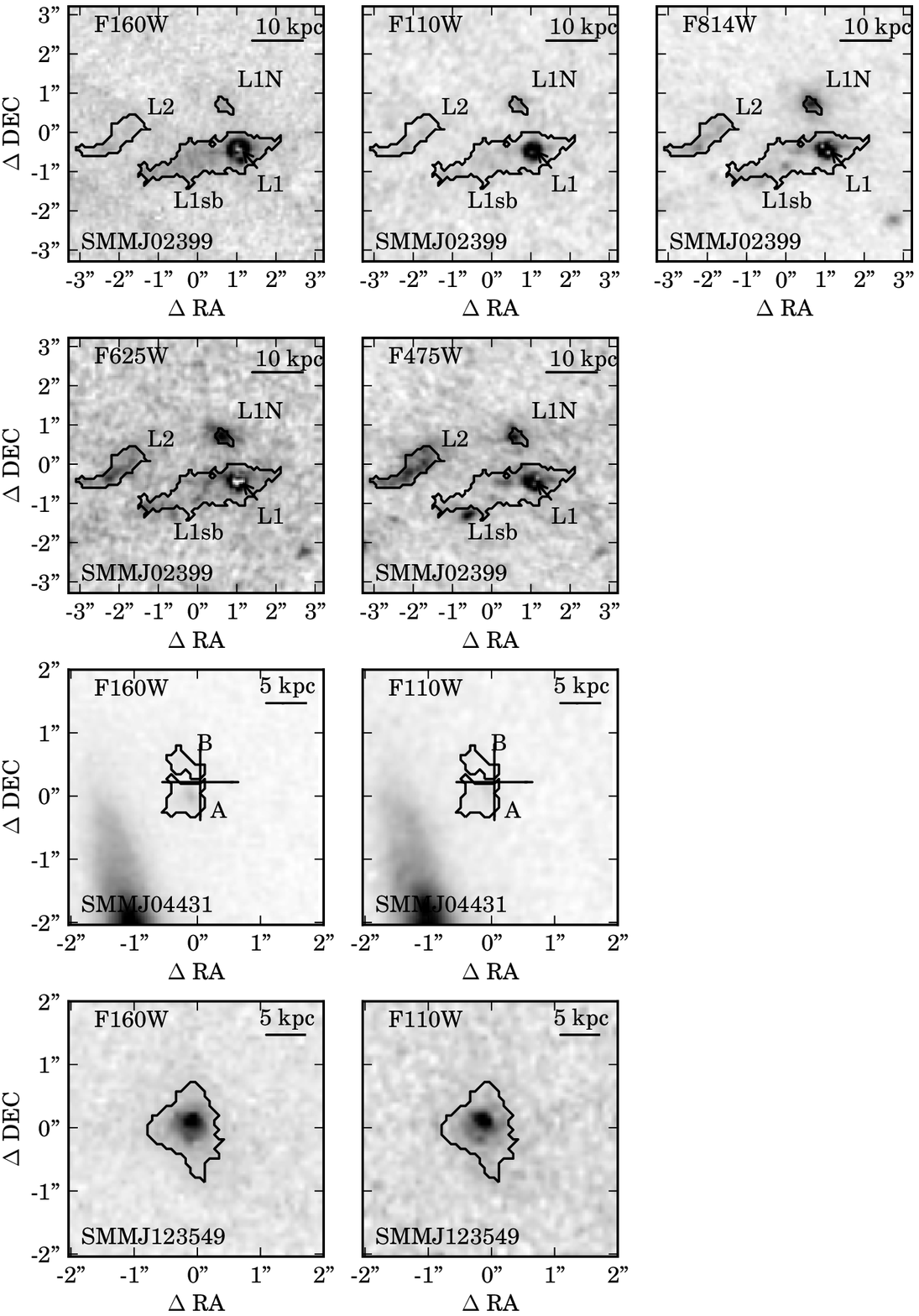}
\caption{  Near-infrared and optical $HST$ images (when available) for our sample of ten SMGs, convolved to match the F160W PSF.  Solid contours represent the boundaries 
of the segmentation regions used for the different components, as described 
in Section 4.3. Axis ticks indicate distance to the source center in 
arcseconds, and the scale bar indicates a reference distance in kpc at the source's redshift, 
before correcting for lensing magnification. North is up and east is to the left. For \Si, the point-source model for the AGN located in L1 has been subtracted from all images. In plots for \Sii, the cross indicates the centroid of the CO(3--2) emission measured by \citet{tacc06} with the total combined position error bars. For \Sv, we show the F850LP image before (fourth panel) and after (third panel) subtraction of the foreground source J1c to demonstrate the results of our GALFIT modelling (see Section \ref{sec:smmj14011_res}). For F110W and F160W we present the final images with J1c removed. \label{fig:stamps}}
\figurenum{1}
\end{figure}

\begin{figure*}
\centering
\includegraphics[scale=0.8]{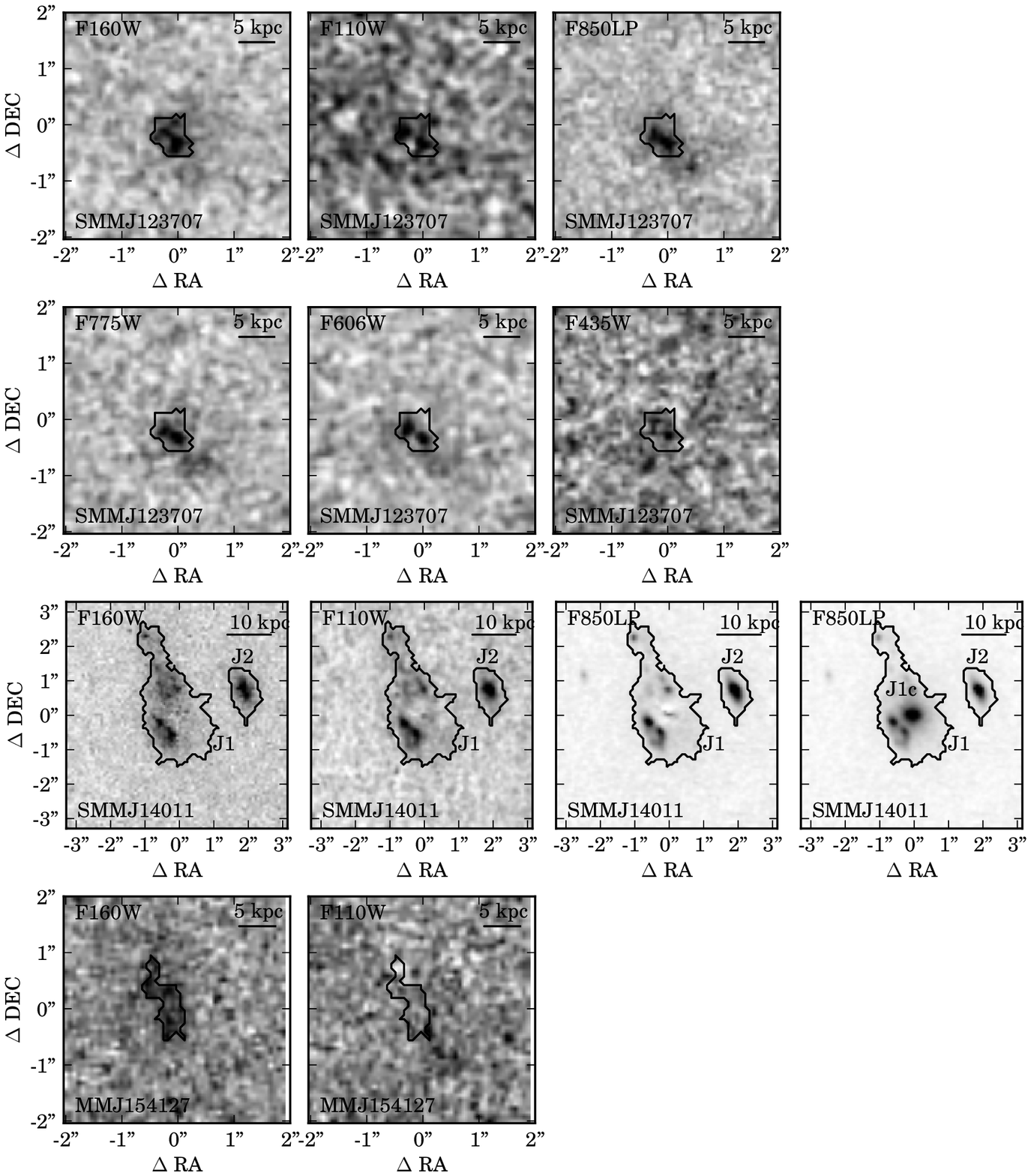}
\caption{ (continued)}
\figurenum{1}
\end{figure*}

\begin{figure*}
\centering
\includegraphics[scale=0.8]{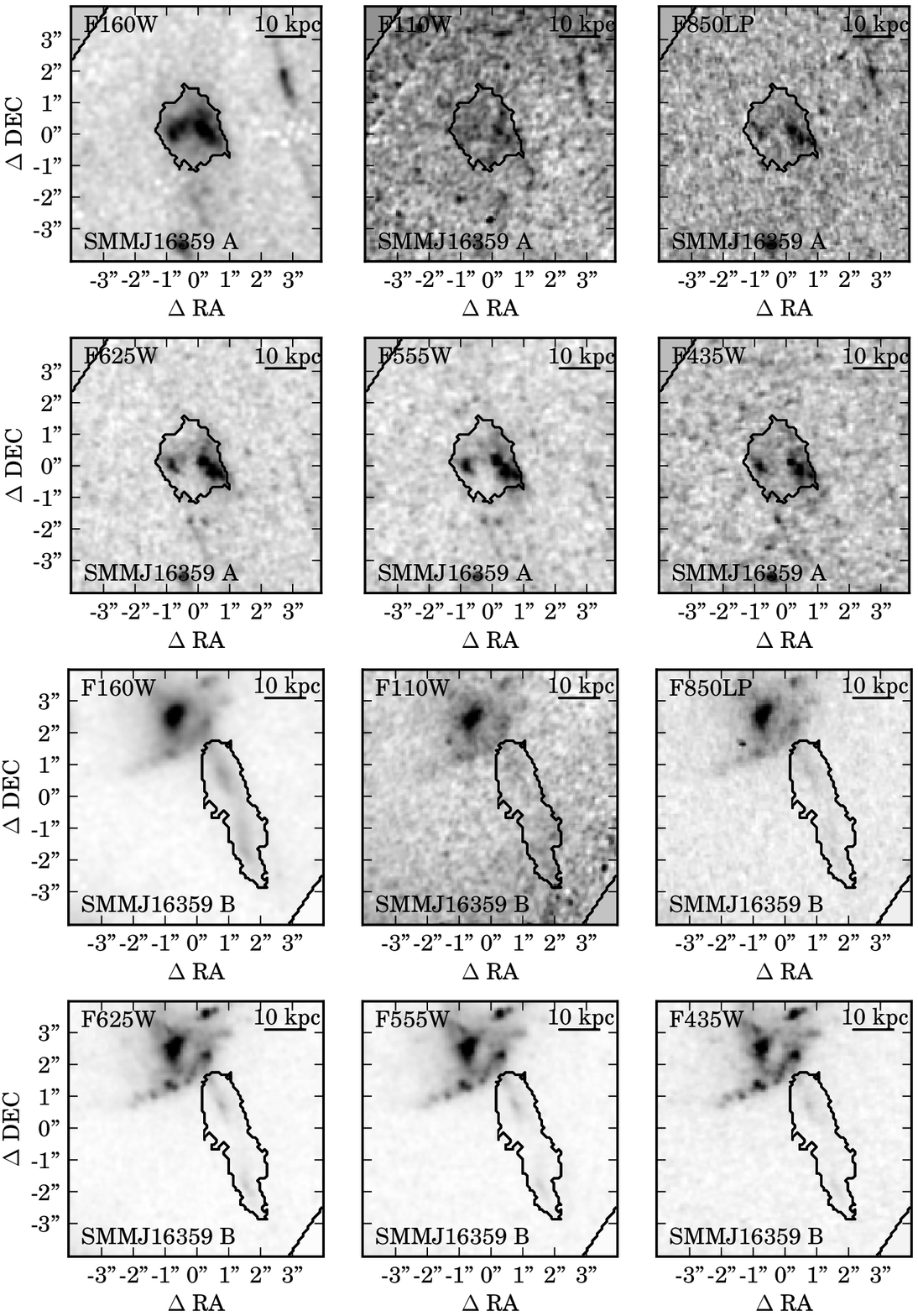}
\caption{ (continued)}
\figurenum{1}
\end{figure*}

\begin{figure*}[ht]
\centering
\includegraphics[scale=0.8]{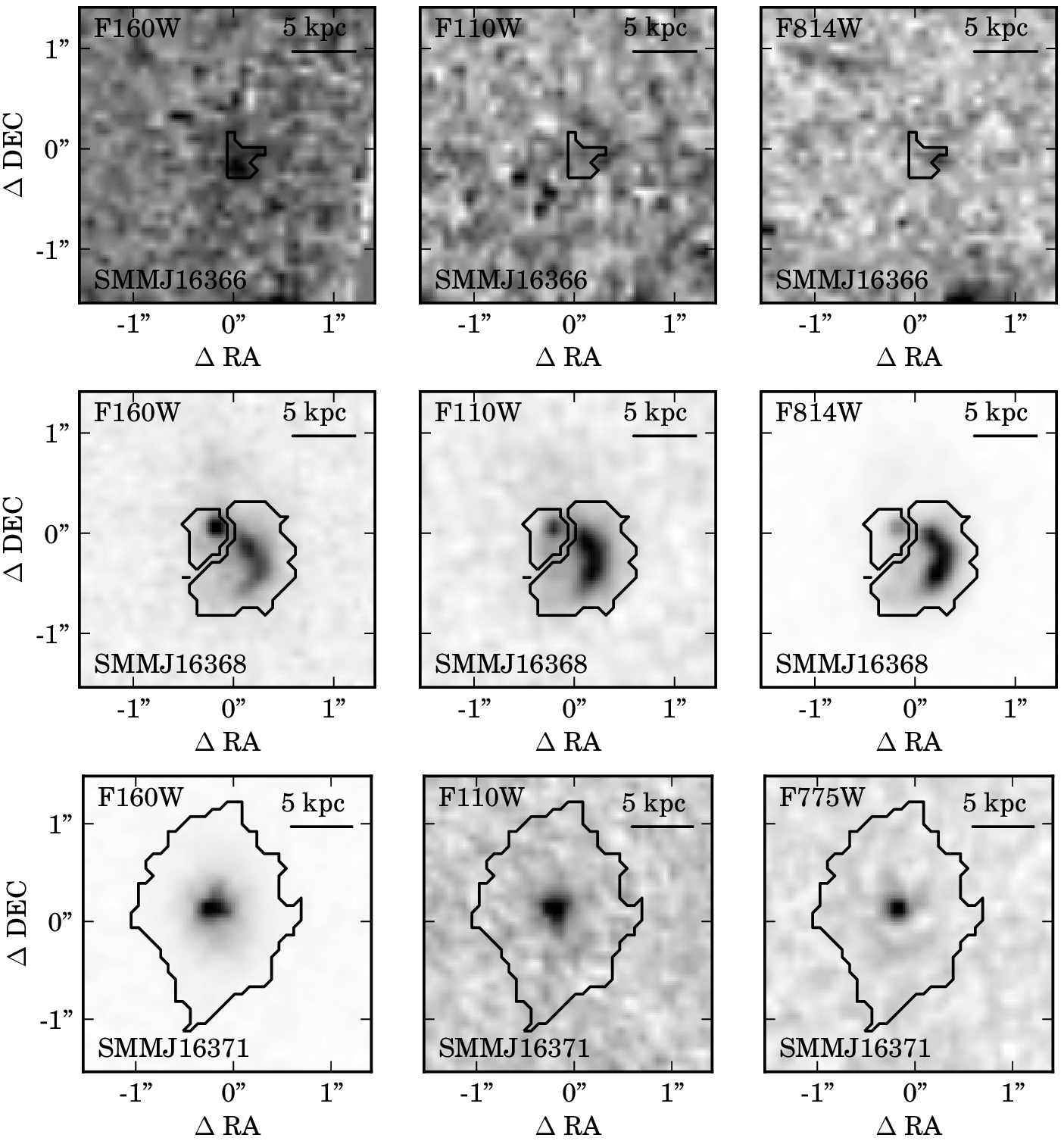}
\caption{ (continued) }
\label{fig:smmj02399}
\figurenum{1}
\end{figure*}

\begin{figure}[ht]
\centering
\figurenum{2}
\includegraphics[scale=0.9]{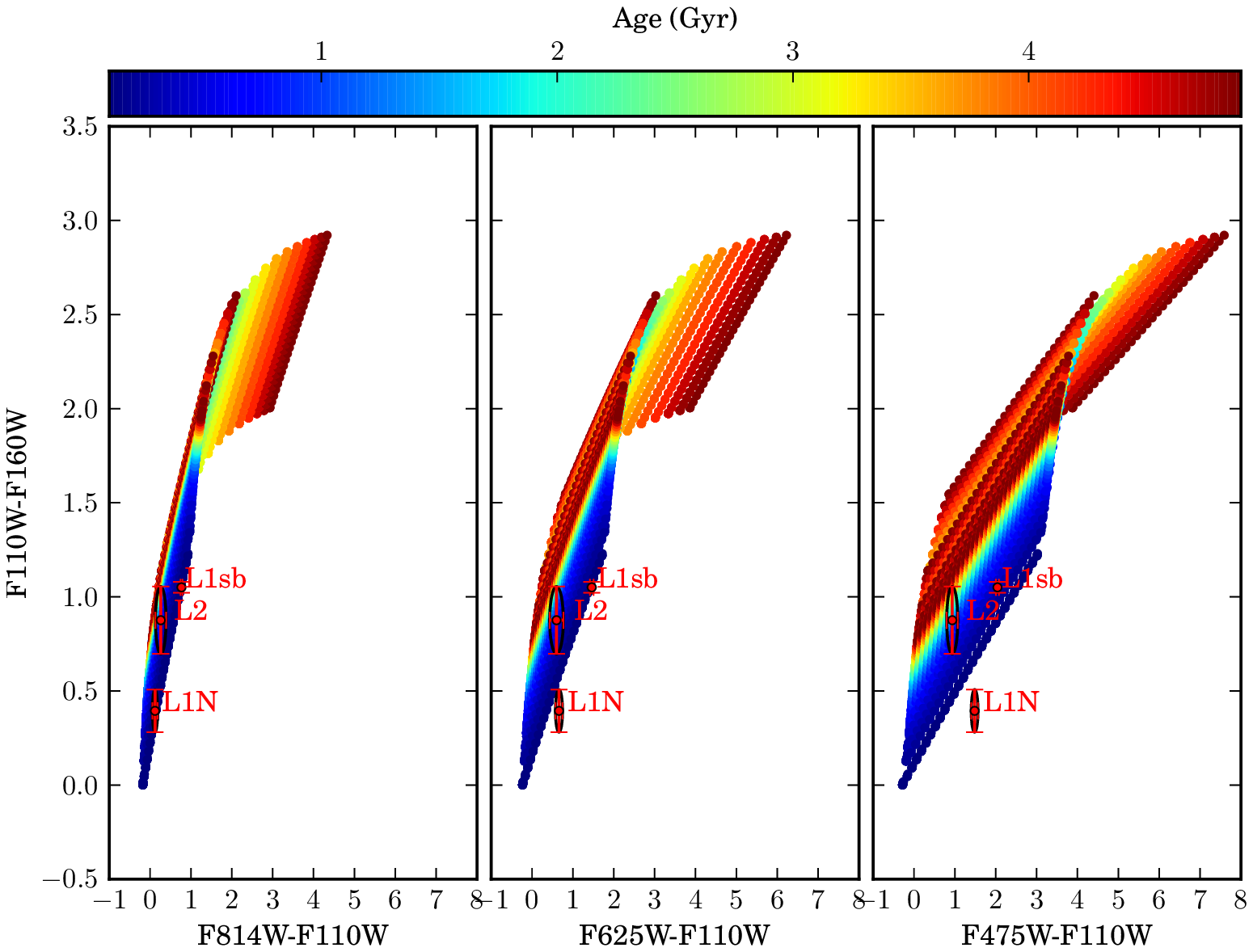}
\caption{Color$-$color plots at the redshift of \Si, for stellar population 
models with exponential star formation timescales between 0 and 5\,Gyr and 
visual extinctions between 0 and 5 mag. Color bar represents the age in 
Gyr; error bars represent the $1\sigma$ photometric errors, and ellipses 
indicate the locations of all stellar population models that are consistent 
within $1\sigma$ with the magnitudes measured for each 
component of \Si.  A color version of this figure is available in the online journal.}
\label{fig:smmj02399_ssp}

\end{figure}

\begin{figure}[ht]
\figurenum{3}
\includegraphics[scale=0.8]{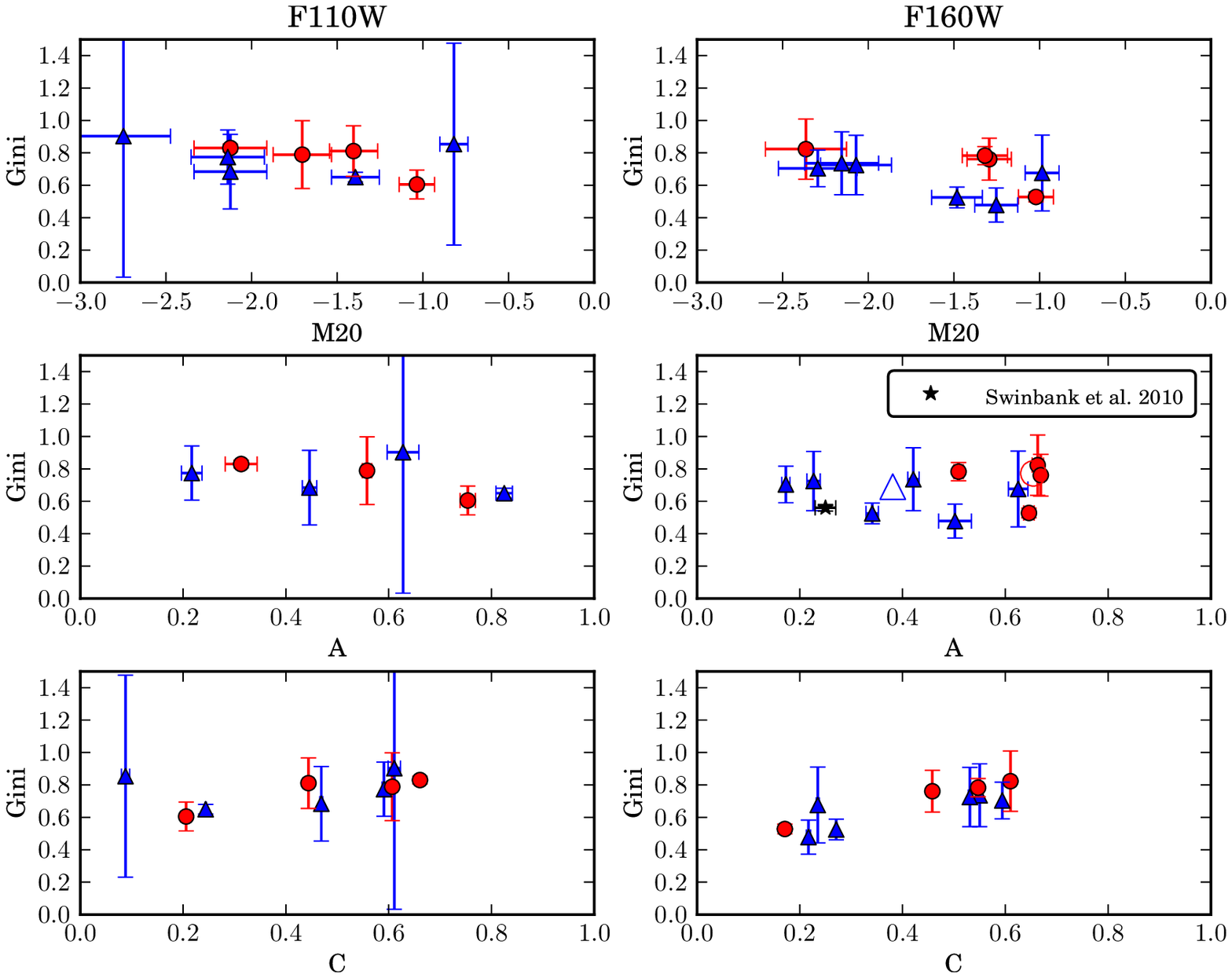}
\caption{Morphological parameters measured in bands F160W and F110W for all 
targets, with their associated uncertainties. Dots correspond to targets 
classified as multi-component objects, and triangles represent 
single-component objects. In the center-right panel, we also plot the median 
value measured by \citet{swin10b} as a star, and the median values for multi-component and single-component objects as an open dot and triangle, respectively.}
\label{fig:morph}

\end{figure}

\clearpage


\begin{center}
\begin{deluxetable}{cccccccc}
\tablecolumns{7}
\tablewidth{0pt}
\tabletypesize{\scriptsize}
\tablecaption{ Near-infrared Observations. \label{tab:obs}}
\tablehead{	
\colhead{} & \colhead{R.A.} & \colhead{decl.}  &  \colhead{} &\multicolumn{2}{c}{Instrument} & \multicolumn{2}{c}{Integration time (s)} \\ 
\cline{5-8}
\colhead{Target }          &   \colhead{(J2000)}   &\colhead{(J2000)}      &  \colhead{$z$} &\colhead{F110W}   &\colhead{F160W} &\colhead{F110W}   &\colhead{F160W}  }
\startdata
\Si	    		& 02:39:52.00	&$-$01:35:59.00 & 2.81 &\nic                    &\nic  			&2560		&2560		\\	 
\Sii	    		& 04:43:07.10	&$+$02:10:25.00 & 2.51 &\nic			&\nic 			&2560		&2560		\\
\Siii	    		& 12:35:49.40	&$+$62:15:37.00 & 2.20 &\nic			&\nic			&2688		&2688		\\
\Siv	    		& 12:37:07.20	&$+$62:14:08.00 & 2.49 &\nic			&\nic  	 		&2688		&2688		\\
\Sv	    		& 14:01:04.90	&$+$02:52:24.00	& 2.56 &\nic			& \nic 			&2560		&2560		\\
\Sx   			& 15:41:27.80	&$+$66:16:17.00 & 2.79 &\nic			& \nic  		&2688		&2688		\\
\Svi	    		& 16:35:54.48	&$+$66:12:30.50 & 2.52 &\nic			& \wfc 			&2688		&2612		\\
\Sviii	    		& 16:36:50.40	&$+$40:57:34.00	& 2.38 &\nic			& \nic\tablenotemark{a}	&2560		&2303 \\		
\Svii	    		& 16:36:58.20	&$+$41:05:24.00 & 2.45 &\nic			&  \nic\tablenotemark{a}&2560		&2303	 \\
\Six   		& 16:37:06.50	&$+$40:53:14.00 & 2.38 &\nic			& \wfc   		&2560		&2412		\\
\enddata
\tablenotetext{a}{Data retrieved from \emph{HST} archive (PID:9856).}
\end{deluxetable}
\end{center}


\begin{center}
\begin{deluxetable}{clcc}
\tablecolumns{7}
\tablewidth{0pt}
\tablecaption{ Optical Observations \label{tab:optical}}
\tablehead{	
\colhead{Target} & \colhead{Filter}  & \colhead{Integration Time (s)}  &\colhead{PID} } 
\startdata
\Si	    		& F475W  & 6780  &11507	\\	 
			&F625W  &  2040&11507\\
			&F814W& 3840 &11507\\ 
\Siv			&F435W &	79200	&9583 \\
			&F606W & 55280 &9583 \\
			&F775W & 87950 &9583 \\
			&F850LP & 254720 &9583 \\
\Sv	    		&F850LP& 9110 & 10154	\\
\Svi	    		& F435W	&5640 & 9717 \\
			&F555W &5640 & 9717 \\
			&F625W& 5640 & 9717\\
			&F850LP& 2680  &9292\\

\Sviii	    		&F814W  & 4760 & 9761\\		
\Svii	    		&F814W  & 4284 & 9761\\
\Six   		&F775W& 2064 & 9856	\\
\enddata
\end{deluxetable}
\end{center}


\clearpage

\begin{center}
\begin{deluxetable}{ccc||ccc}
\tablecolumns{9}
\tablewidth{0pt}
\tabletypesize{\scriptsize}
\tablecaption{ Photometry Results.  \label{tab:colors} }
\tablehead{
\colhead{Object}   & \colhead{Filter} & \colhead{$m_{\rm AB}$} & \colhead{Object}   & \colhead{Filter} & \colhead{$m_{\rm AB}$}\\  }
\startdata

\Si\	  L1N   & F160W	& 25.02 $\pm$ 0.07  &	 	\Sx   	        & F160W  &23.79 $\pm$ 0.05 \\
		 & F110W	&  25.46 $\pm$ 0.04 &			        & F110W	 & $\geq25.70$   \\
               	 &F814W	&   25.56 $\pm$ 0.02 &   	            \Svi\ A	  & F160W	 & 25.03 $\pm$ 0.02 \\
	           &F625W	&   26.10 $\pm$ 0.04 &			  & F110W	 & 26.01 $\pm$ 0.08 \\ 	
		 &F475W	&  26.92 $\pm$ 0.04&	 		 & F850LP       & 26.10 $\pm$ 0.07 	 \\
 \Si	\  L2\  & F160W	& 24.45 $\pm$ 0.06 &                    	& F625W	 & 26.28 $\pm$ 0.02 \\
                     & F110W	& 25.01 $\pm$ 0.04 &                    		& F555W	 & 26.51 $\pm$ 0.03  \\	 
                     & F814W	& 25.25 $\pm$ 0.02 &				& F435W	 & 27.03 $\pm$ 0.04  \\
                     &F625W	& 25.57 $\pm$ 0.03&	\Svi\ B		 & F160W	&25.08 $\pm$ 0.01\\
	 	 & F475W	& 25.64 $\pm$ 0.03&				 & F110W	&25.69  $\pm$ 0.05 \\ 
\Si\	  L1sb  & F160W	& 21.85  $\pm$	0.01	&   			& F850LP	&26.08 $\pm$ 0.04\\
		 & F110W	&  22.89 $\pm$ 0.02	&			& F625W	& 26.37 $\pm$ 0.01  \\
		 &F814W	&  23.63 $\pm$ 0.01 	&			&F555W	&26.53 $\pm$ 0.02 \\
		 &F625W	&  24.31 $\pm$ 0.03	&			& F435W	&27.04 $\pm$ 0.03 \\
		 &F475W	&  24.89 $\pm$ 0.02	&		\Sviii	\  B 		&F160W	& 21.88 $\pm$ 0.01 \\
\Sii\  A   	 &  F160W &  24.72 $\pm$ 0.04  &					&F110W	&22.63 $\pm$ 0.01\\				
		  & F110W  & 25.83 $\pm$ 0.07   &					&F814W	&22.64 $\pm$ 0.00 \\
\Sii\  B    	&  F160W &  25.83 $\pm$ 0.07 &		\Sviii	\  C   &F160W	& 22.92 $\pm$ 0.02  \\  
		& F110W  & 26.94 $\pm$ 0.13  &				&F110W	&24.06 $\pm$ 0.02 \\  
\Siii	    	& F160W  &  21.92 $\pm$ 0.01 &				&F814W	&22.63 $\pm$ 0.01 \\
		& F110W & 22.80 $\pm$ 0.02   &		\Svii	    	& F160W    &24.41 $\pm$ 0.08 \\
 \Siv	    	& F160W   & 23.30 $\pm$ 0.03  & 				& F110W	& $\geq25.58$ \\
		& F110W	& 25.08 $\pm$0.07 &  				&F814W	&  $\geq27.17$  \\
											
		&F850LP  &     $25.57\pm 0.04$&   			\Six   	&F160W	&21.93 $\pm$ 0.01\\
				&F775W  &	 $26.15 \pm 0.05$ &   	   	&F110W	&23.66 $\pm$ 0.07 \\
		&F606W  &	 $26.85 \pm 0.06$&   				&F775W	&24.40 $\pm$ 0.06  \\
		&F435W  &   $\geq27.73$ 	&   				&	&  \\
\Sv\	   J1  	& F160W    &  22.58 $\pm$ 0.08   &   	&	& \\
		& F110W	& 23.41 $\pm$ 0.09   &   	&	& \\
		& F850LP    & 23.56 $\pm$ 0.05  &   	&	& \\
\Sv\  J2 	& F160W    & 23.92 $\pm$ 0.10  &   	&	& \\
		& F110W	& 24.66 $\pm$ 0.11	  &   	&	& \\
		&F850LP 	&	24.49 $\pm$ 0.06  &   	&	& \\
	
\enddata
\tablecomments{We report the magnitudes or limits measured inside the segmentation regions defined as described in Section 4.3. All magnitudes are corrected for gravitational magnification using the amplification factors quoted in the text.}
\end{deluxetable}
\end{center}


\clearpage

\begin{center} 
\begin{deluxetable}{ccccccccccccc} 
\tablecolumns{13} 
\tablewidth{0pt} 
\tabletypesize{\scriptsize} 
\tablecaption{ Morphological Parameters Calculated from F160W and F110W Images of Our Targets. \label{tab:morph}} 
\tablehead{\colhead{}  &\colhead{}& \multicolumn{5}{c}{F160W} &  \multicolumn{5}{c}{F110W}  \\ 
 \colhead{Object}  &  \colhead{$\mu$}  & \colhead{$r_{\rm{h}} $(kpc)}  &\colhead{$A$}   & \colhead{$C$}  &\colhead{$G$}  & \colhead{$M_{20}$}  &\colhead{$r_{\rm{h}}$ (kpc)} &\colhead{$A$}&   \colhead{$C$}   & \colhead{$G$}  &\colhead{$M_{20}$}  & \colhead{S/M}  } 
\startdata
SMM\,J02399$-$0136 & 2.4 & 3.89 & 0.66 & 0.61 & 0.82 & -2.36    & 1.65 & 0.31 &0.66 & 0.83 & -2.12 &M \\
SMM\,J04431+0210 & 4.4 & 3.67 & 0.65 & 0.17 & 0.53 & -1.02    & 3.63 & 0.75 &0.21 & 0.60 & -1.03 & M \\

SMM\,J123549+621536 & \nodata & 3.21 & 0.42 & 0.55 & 0.74 & -2.15    & 3.66 & 0.45 &0.47 & 0.68 & -2.12 & S\\
SMM\,J123707+621410 & \nodata & 2.78 & 0.34 & 0.27 & 0.53 & -1.48    & 3.19 & 0.82 &0.24 & 0.65 & -1.39 & S\\
SMM\,J14011+0252 J1 & 4.0 & 8.36 & 0.67 & 0.46 & 0.76 & -1.29    & 7.95 & 1.04 &0.44 & 0.81 & -1.41 & M\\
SMM\,J14011+0252 J2 & 3.5 & 2.82 & 0.23 & 0.53 & 0.72 & -2.07    & 1.89 & 0.22& 0.59 & 0.77 & -2.14 & M\\

MM\,J154127+6616 & \nodata & 3.55 & 0.62 & 0.23 & 0.68 & -0.98     & \nodata        & \nodata    & \nodata     &\nodata     &\nodata& S \\
SMM\,J16368+4057 & \nodata & 2.76 & 0.51 & 0.55 & 0.78 & -1.32    & 3.07 & 0.56 &0.61 & 0.79 & -1.70 & M\\
SMM\,J16366+4105 & \nodata & 1.81 & 0.38 & 0.22 & 0.48 & -1.25    & \nodata         & \nodata    & \nodata    &\nodata     &\nodata& S \\

SMM\,J16371+4053 & \nodata & 3.36 & 0.17 & 0.59 & 0.70 & -2.29    & 2.30 & 0.63 &0.61 & 0.90 & -2.75 & S\\

\enddata
\tablecomments{Objects \Sx\ and \Svii\ were not detected in F110W, so their morphologies were only studied in F160W. In the second column we indicate the magnification factors $\mu$ assumed for lensed objects; $r_{\rm{h}}$ corresponds to the circularized radius, calculated as $r_{\rm{h}}=\sqrt{a_{\rm{h}}b_{\rm{h}}}$, where $a_{\rm{h}}$ and $b_{\rm{h}}$ are the major and minor half light radii measured from the images. The last column indicates whether the object  was classified as a single (S) or multiple (M) component system. SMM\.J16359+6612 is excluded from the morphological analysis due to its strong distortion by the gravitational potential.}
\end{deluxetable}
\end{center}


\clearpage
\begin{center}
\begin{deluxetable}{lccc}
\tablecolumns{4} 
\tablewidth{0pt} 
\tablecaption{ Stellar Masses, Derived Using a \citet{chab03} IMF. \label{tab:mass}}
 \tablehead{\colhead{Object}  &  \colhead{$\log_{10}(M_*/M_{\odot})$} & \colhead{Mass ratio} & \colhead{S/M} } 
\startdata
SMM\,J02399$-$0136 L1sb & $11.8\pm 0.1$  &  & M\\
SMM\,J02399$-$0136 L2 & $10.1\pm 0.1$ &  $50.1 $ & \\
SMM\,J02399$-$0136 L1N & $ \,\,9.7\pm 0.1$ & $125.8$ & \\ \hline

SMM\,J04431+0210  A & $10.0\pm 0.3$ & & M  \\
SMM\,J04431+0210  B & $ \,\,9.6 \pm 0.4$ & $2.5$ & \\\hline

SMM\,J123549+621536 & $10.8 \pm 0.3$ & \nodata & S\\\hline
SMM\,J123707+621410 & $11.3\pm 0.1$&\nodata  & S  \\\hline

SMM\,J14011+0252 J1 & $10.5\pm 0.2$ &  & M\\
SMM\,J14011+0252 J2& $9.8\pm 0.3$ & $5.0$ & \\ \hline

MM\,J154127+6616 & \nodata & \nodata & S \\\hline
SMM\,J16359+6612 A & $ \,\,9.6\pm 0.2$ &\nodata & S\\\hline

SMM\,J16368+4057 B &  $10.6\pm 0.2$ &     & M\\
SMM\,J16368+4057 C & $10.7\pm 0.2$ &  1.3  & \\\hline

SMM\,J16366+4105 & $11.8\pm0.2$ &  \nodata & S\\\hline
SMM\,J16371+4053 & $11.7\pm0.1$ &  \nodata & S\\ 
\enddata
\tablecomments{ In the second column we report the stellar masses of the individual components identified in multiple systems, and in the third column we give the mass ratio. The third column is repeated from Table \ref{tab:morph}, and indicates whether an object was classified as a single (S) or multiple (M) system.}
\end{deluxetable}
\end{center}

\end{document}